# DESIGN OF AN INDUCTION PROBE FOR SIMULTANEOUS MEASUREMENTS OF PERMITTIVITY AND RESISTIVITY

# PROGETTAZIONE DI UNA SONDA AD INDUZIONE PER MISURE SIMULATANEE DI PERMITTIVITÁ E RESISITIVITÁ

**(Short Title - Titolo Abbreviato: RESPER *PROBE*)**

Alessandro Settimi, Achille Zirizzotti, James A. Baskaradas, Cesidio Bianchi

*INGV (Istituto Nazionale di Geofisica e Vulcanologia) – via di Vigna Murata 605, I-00143 Rome, Italy*

# Index

## Abstract

In this paper, we propose a discussion of the theoretical design and move towards the development and engineering of an induction probe for electrical spectroscopy which performs simultaneous and non invasive measurements on the electrical RESistivity $\rho$ and dielectric PERmittivity $\varepsilon_r$ of non-saturated terrestrial ground and concretes (RESPER probe). In order to design a RESPER which measures $\rho$ and $\varepsilon_r$ with inaccuracies below a prefixed limit (*10%*) in a band of low frequencies (LF) (*B=100kHz)*, the probe should be connected to an appropriate analogical digital converter (ADC), which samples in uniform or in phase and quadrature (IQ) mode, otherwise to a lock-in amplifier. The paper develops only a suitable number of numerical simulations, using Mathcad, which provide the working frequencies, the electrode-electrode distance and the optimization of the height above ground minimizing the inaccuracies of the RESPER, in galvanic or capacitive contact with terrestrial soils or concretes, of low or high resistivity. As findings of simulations, we underline that the performances of a lock-in amplifier are preferable even when compared to an IQ sampling ADC with high resolution, under the same operating conditions. As consequences in the practical applications: if the probe is connected to a data acquisition system (DAS) as an uniform or an IQ sampler, then it could be commercialized for companies of building and road paving, being employable for analyzing "in situ" only concretes; otherwise, if the DAS is a lock-in amplifier, the marketing would be for companies of geophysical prospecting, involved to analyze "in situ" even terrestrial soils.

## Riassunto

In questo articolo, proponiamo una discussione del progetto teorico e ci muoviamo verso lo sviluppo e l'ingegnerizzazione di una sonda ad induzione per la spettroscopia elettrica che effettui misure simultanee e non invasive delle RESistività elettrica $\rho$ e PERmittività dielettrica $\varepsilon_r$ per suoli terrestri e calcestruzzi non saturi (sonda RESPER). Al fine di progettare un RESPER che misuri $\rho$ e $\varepsilon_r$ con incertezze al di sotto di un limite prefissato (*10%*) in una banda di basse frequenze (*LF*) (*B=100kHz*), la sonda dovrebbe essere connessa ad un convertitore analogico digitale (*ADC*) adeguato, che campioni in modo uniforme o in fase e quadratura (IQ), altrimenti ad un amplificatore *lock-in*. Il lavoro sviluppa esclusivamente un opportuno numero di simulazioni numeriche che, usando Mathcad, forniscono le frequenze di lavoro, la distanza elettrodo-elettrodo e l'ottimizzazione dell'altezza da terra che riducono al minimo le incertezze del RESPER, a contatto galvanico o capacitivo con suoli terrestri o calcestruzzi, di resistività bassa o alta. Come risultati delle simulazioni, teniamo a sottolineare che le prestazioni di un amplificatore *lock-in* sono preferibili anche ad un ADC a campionamento IQ con alta risoluzione di bit, sotto le stesse condizioni di funzionamento. Come conseguenze nelle applicazioni pratiche: se la sonda è connessa ad un sistema di acquisizione dati (*DAS*), come una campionatore uniforme o IQ, allora potrebbe essere commercializzata per le imprese di costruzione e di pavimentazione stradale, ed utilizzabile per l'analisi "in situ" solo dei calcestruzzi, altrimenti, se il *DAS* è un amplificatore *lock-in*, la commercializzazione sarebbe per le società di prospezione geofisica, interessate ad analizzare "in situ" anche i suoli terrestri.

## Introduction.

Electrical resistivity and dielectric permittivity are two independent physical properties which characterize the behavior of bodies when these are excited by an electromagnetic field. The measurements of these properties provides crucial information regarding practical uses of bodies (for example, materials that conduct electricity) and for countless other purposes.

We refer to [Grard, 1990, a-b][Grard and Tabbagh, 1991][Tabbagh et al.,1993], who have verified experimentally that the complex permittivity (resistivity and dielectric constant) of ground can be measured with a set of four electrodes carried on a vehicle. This new approach offers significant advantages over the system first introduced by Wenner which gives only the resistivity and requires the manual insertion of the four electrodes into the ground.

This technique allows for relatively high resistivity ranges where electromagnetic induction methods are not applicable because the resistivity effect is too low and hidden by the magnetic susceptibility response. The determination of the dielectric constant, which previously was exclusively based on high-frequency radar techniques (above *30MHz*), is now possible in a wide range of lower frequencies, which is extremely important because this parameter is very sensitive to the presence of water.

Specially, we refer to [Vannaroni et al., 2004][Del Vento and Vannaroni, 2005]. The soil dielectric spectroscopy probe (SDSP) determines the complex permittivity of the shallow subsoil from measurements of the mutual impedance of a four-electrode system electrically coupled with the ground. The mutual impedance is defined as the ratio of the voltage measured across a pair of receiving electrodes to the current transmitted by a second pair of electrodes. The mutual impedance depends on the geometry of the electrode array but also on the complex permittivity of the ground.

The measurements are performed in AC regime and the electrodes do not require low impedance DC contacts with the soil. In fact, the terminals are capacitively coupled to the terrain and do not require any galvanic contact. This feature presents two advantages. The first is that one can inject into the ground the exciter current even in absence of effective electrodes-soil galvanic contact, making the system particularly suitable to be hosted onboard a moving vehicle. The second relies on the fact that, in AC regime, one can measure not only the conduction but also the displacements currents in the ground, thus obtaining information on the polarizability of the substances contained in it.

The band of frequency is limited to the range *10kHz-1MHz*. The lower limit is effectively imposed by two facts: a) firstly, the Maxwell-Wagner effect which limits probe accuracy [Frolich, 1990], as the most important limitation happens because of interface polarization effects that are stronger at low frequencies, say below *1kHz* depending of medium conductivity; b) secondly, the need to maintain the amplitude of the current at measurable levels as, given the capacitive coupling between electrodes and soil, the current magnitude is proportional to the frequency. On the other hand, the upper limit is suitably fixed to allow the analysis of the system in a regime of quasi static approximation and neglect the velocity factor of the cables used for the electrode harness, that in turn degrades the accuracy of the mutual impedance phase measurements.

## Topic and structure of the paper.

In the present paper, we propose a discussion of the theoretical design and move towards the development and engineering of a induction probe for electrical spectroscopy which acquires complex impedance in the field. By increasing the distance between the electrodes, it is possible to investigate the electrical properties of sub-surface structures to greater depth. The probe can perform immediate measurements on materials with high resistivity and permittivity, without subsequent stages of data analysis.

We refer to our previous papers [Settimi et al, 2009, a-b]. The first paper [Settimi et al, 2009, a] had discussed the theoretical modelling of an induction probe which performs simultaneous and non invasive measurements on the electrical RESistivity $\rho$ and dielectric PERmittivity $\varepsilon_r$ of non-saturated media (RESPER probe). A mathematical-physical model was applied on propagation of errors in the measurement of resistivity and permittivity based on the sensitivity functions tool [Murray-Smith, 1987]. The findings were also compared to the results of the classical method of analysis in the frequency domain, which is useful for determining the behaviour of zero and pole frequencies in the linear time invariant (LTI) circuit of the quadrupole. The paper underlined that average values of electrical resistivity and dielectric permittivity may be used to estimate the complex impedance over various terrains [Edwards, 1998] and concretes [Polder et al., 2000][Laurents, 2005], especially when they are characterized by low levels of water content [Knight and Nur, 1987] and analyzed within a bandwidth ranging only from LF to MF frequencies [Myounghak et al., 2007][Al-Qadi et al., 1995]. In order to meet the design specifications which ensure satisfactory performances of the probe (inaccuracy no more than *10%*), the forecasts provided by the sensitivity functions approach were discussed in comparison to those foreseen by the transfer functions method (in terms of both the frequency $f$ and measurable range of resistivity $\rho$ or permittivity $\varepsilon_r$).
The second paper [Settimi et al, 2009, b] moves towards the practical implementation of electrical spectroscopy. In order to design a RESPER which measures $\rho$ and $\varepsilon_r$ with inaccuracies below a prefixed limit (*10%*) in a band of LF (*B=100kHz)*, the probe should be connected to an appropriate analogical digital converter (ADC), which samples in phase and quadrature (IQ) [Jankovic and Öhman, 2001] or in uniform [Razavi, 1995] mode. If the probe is characterized by a galvanic contact with the surface, then the inaccuracies in the measurement of resistivity and permittivity, due to the IQ or uniform sampling ADC, can be analytically expressed. A large number of numerical simulations have proved that the performance of the probe depends on the selected sampler and that the IQ is preferable when compared to the uniform mode under the same operating conditions, i.e. number of bits and medium.

This paper develops only a suitable number of numerical simulations, using Mathcad, which provide the working frequencies, the electrode-electrode distance and the optimization of the height above ground minimizing the inaccuracies of the RESPER, in galvanic or capacitive contact with terrestrial soils or concretes of low or high resistivity. As findings of simulations, we underline that the performances of a lock-in amplifier [Scofield, 1994] are preferable even when compared to an IQ sampling ADC with high resolution, under the same operating conditions. As consequences in the practical applications: if the probe is connected to a DAS as an uniform or an IQ sampler, then it could be commercialized for companies of building and road paving, being employable for analyzing “in situ” only concretes; otherwise, if the DAS is a lock-in amplifier, the marketing would be for companies of geophysical prospecting, involved to analyze “in situ” even terrestrial soils.
The paper is organized as follows. In section 1, Mathcad simulations on non-saturated terrestrial soils and concretes are reported when the RESPER is connected to uniform (1.1) or IQ (1.2) sampling ADCs otherwise to lock-in amplifiers (1.3). Sec. 2 deepens further the discussion on the design for the characteristic geometrical dimensions of the probe. Conclusions are drawn in sec. 3.

## 1. Mathcad simulations on non-saturated terrestrial ground and concretes.

In this section, which refers to [Settimi et al, 2009, a-b], we propose to develop a suitable number of simulations, using Mathcad, to design a RESPER, in galvanic or capacitive contact with a subjacent medium, performing simultaneous and non invasive measurements for the electrical conductivity $\sigma$ and the dielectric permittivity $\varepsilon_r$. The RESPER is connected to a DAS, as an uniform / IQ sampling ADC or an lock-in amplifier, which is appropriate for surfaces as the terrestrial ground or the concretes with high and low $\sigma$. Deepening, the Mathcad simulations are applied to a quadrupolar probe [fig. 1], which interacts with the medium by

- a galvanic contact or
- a capacitive contact,

the quadrupole being connected to a DAS [fig. 2], with specifications only similar to those of

1. one of the uniform sampling ADCs, NI-USB(or PCI, PXI)-51XX(or 61XX), with high speed or high resolution or high density [Razavi, 1995], commercialized by the National Instruments Company, or
2. an IQ sampling ADC [Jankovic and Öhman, 2001], with high or low resolution, which are being designed in our laboratories, or otherwise
3. one of lock-in amplifiers *SR810* and *SR830* [Scofield, 1994], commercialized by the Standford Research Systems Company.

So, the simulations provide the ranges of measurability for the conductivity $\sigma$ and the permittivity $\varepsilon_r$, describing various terrestrial soils and concretes, which minimize the inaccuracies in the measurements of the probe, in galvanic contact, once designed its characteristic geometrical dimensions and selected its frequency band, i.e. $B=100kHz$. Besides, the simulations provide the working frequencies, the electrode-electrode distance and the optimization of the height above ground which minimize the inaccuracies of the probe, in galvanic or capacitive contact with non-saturated terrestrial soils [Edwards, 1998], of high and low conductivity, respectively as

A. flat, marshy, densely wooded in places ($\rho=1/\sigma=130\Omega\cdot m,\ \varepsilon_r=13$) and
B. high-rise city centres, industrial areas ($\rho=3000\Omega\cdot m,\ \varepsilon_r=4$),

or for concretes [Polder et al., 2000][Laurents et al., 2005] as (indoor climate carbonated) blast fumace slag, fly ash cement and silica concrete fume, respectively with

C. high conductivity ($\rho=1/\sigma=4000\Omega\cdot m,\ \varepsilon_r=9$) and
D. low conductivity ($\rho=10000\Omega\cdot m,\ \varepsilon_r=4$).

Even if, according to Debye polarization mechanisms [Debye, 1929] or Cole-Cole diagrams [Auty and Cole, 1952], the complex permittivity of various materials in the frequency band from very low (VLF) to very high (VHF) frequencies exhibits several intensive relaxation effects and a non-trivial dependence on the water saturation [Chelidze and Gueguen, 1999][Chelidze et al., 1999], anyway average values of electrical resistivity and dielectric permittivity may be used to estimate the complex impedance over various terrains and concretes, especially when they are characterized by low levels of water content and analyzed within a frequency bandwidth ranging only from LF to MF frequencies.

When $h=0$, i.e. the RESPER exhibits a galvanic contact with a subjacent medium characterized by an electrical conductivity $\sigma$ and a dielectric permittivity $\varepsilon_r$, the transfer impedance is a function of the working frequency $f$ such that its modulus $Z_N(f)$ is constant down to the cut-off frequency $f_T=f_T(\sigma,\varepsilon_r)=\sigma/(2\pi\varepsilon_0(\varepsilon_r+1))$. For a terrestrial ground of type A, $f_T=9.876MHz$; type B, $f_T=1.198MHz$; for concretes of type C, a lower value $f_T=449.378kHz$, as the ratio $\sigma/\varepsilon_r$ is lower in the concretes than in the terrestrial soils; type D close to type C, $f_T=359.502kHz$, as $\sigma/\varepsilon_r$ changes not so much in the concretes C and D. Only if the quadrupole probe is in galvanic contact with the subjacent medium, i.e. $h=0$, then our mathematical-physical model predicts that the inaccuracies $\Delta\sigma/\sigma(f)$ for $\sigma$ and $\Delta\varepsilon_r/\varepsilon_r(f)$ for $\varepsilon_r$ are invariant in the linear (Wenner's) or square configuration and independent from the characteristic geometrical dimension of the quadrupole, i.e. electrode-electrode distance $L$ [Settimi et al., 2009, a].
When the quadrupole is in capacitive contact with the medium, an optimal ratio $x_{opt}=h_{pot}/L$ should be fixed between the height $h_{opt}$ above ground and $L$, in order to perform optimal measurements of the permittivity $\varepsilon_r$. The square [fig. 3.a] configuration should be designed with an optimal height/dimension ratio $x_{opt,S}$ slightly smaller than the corresponding linear (Wenner's) [fig. 3.b] ratio $x_{opt,W}$. In fact, for a terrestrial ground of type A, $x_{opt,S}=0.046$ and $x_{opt,W}=0.052$; type B, $x_{opt,S}=0.078$ and $x_{opt,W}=0.087$; for concretes of type C, $x_{opt,S}=0.055$ and $x_{opt,W}=0.062$; type D identical to type B, $x_{opt,S}=0.078$ and $x_{opt,W}=0.087$, due to an equal permittivity $\varepsilon_r=4$. Then, both the configurations provide, in correspondence of the two ratio, i.e. $x_{opt,S}$ and $x_{opt,W}$, almost an invariant position of the zero and pole frequencies for the transfer impedance in modulus, i.e. $z(x_{opt})$ and $p(x_{opt})$. Thus, for terrestrial soils of type A, $z_A=3.878MHz$ and $p_A=15.514MHz$; type B, $z_B=470.587kHz$ and $p_B=1.882MHz$; for concretes of type C, $z_C=176.47kHz$ and $p_C=705.881kHz$; type D, similar to type C, $z_D=141.176kHz$ and $p_D=564.705kHz$, as $\sigma/\varepsilon_r$ is almost constant in the concretes C and D. Besides, chosen the DAS, both the configurations provide, in correspondence of the two ratio, i.e. $x_{opt,S}$ and $x_{opt,W}$, almost identical values for the inaccuracies in the measurements of $\sigma$ and $\varepsilon_r$, i.e. $\Delta\sigma/\sigma(f)$ and $\Delta\varepsilon_r/\varepsilon_r(f)$ [Settimi et al, 2009, a].

### 1.1. RESPER connected to uniform sampling ADCs.

If the RESPER, of frequency band $B=100kHz$, has a galvanic contact with the subjacent medium, i.e. $h=0$, then, in order to measure the electrical conductivity $\sigma$ and dielectric permittivity $\varepsilon_r$ with inaccuracies, respectively $\Delta\sigma/\sigma$ and $\Delta\varepsilon_r/\varepsilon_r$, below a prefixed limit ($10\%$), an uniform sampling ADC does not allow to perform measurements on terrestrial soils characterized by an high conductivity $\sigma$ (type A) [fig. 4.a]. In fact, it is necessary $n_{min}=18$ [tab. 1.a] as minimal limit for the number of bits, too high to render the measuring system insensitive to the electrical noise of the external environment, as discussed in ref. (Settimi et al, 2009, b). Instead, the uniform sampling ADCs guarantee measurement of $\varepsilon_r$ on terrestrial soils with a low $\sigma$ (type B) [fig. 4.b]. In fact, it is sufficient to have $n_{min}=12$, even if the minimum value of working frequency $f_{min}$ is too close to the end of the band $B$. In the best case, i.e. a high speed ADC only similar to the *NI PCI(PXI)-5124*, with sampling rate $f_S=200MHz$, it results $f_{min}=95.055kHz$, such that $\Delta\sigma/\sigma(f_{min})=2.4\cdot 10^{-3}$ [besides $f_{opt}=387.772kHz$, where $\Delta\sigma/\sigma(f_{opt})=8.825\cdot 10^{-3}$, $\Delta\varepsilon_r/\varepsilon_r(f_{opt})=0.017$] [tab. 1.b] [Settimi et al, 2009, b].
Moreover, an uniform sampling ADC guarantees to perform measurements on concretes characterized by an high conductivity $\sigma$ (type C) [fig. 4.c], with $n_{min}=12$ as minimal number of bits, though working in a small band $[f_{min}, B]$. In the best case, again a high speed ADC similar to the *NI PCI(PXI)-5124*, the minimum value of frequency is $f_{min}=33.315kHz$, such that $\Delta\sigma/\sigma(f_{min})=1.16\cdot 10^{-3}$ [besides $f_{opt}=193.94kHz$, where $\Delta\sigma/\sigma(f_{opt})=4.923\cdot 10^{-3}$, $\Delta\varepsilon_r/\varepsilon_r(f_{opt})= 8.282\cdot 10^{-3}$] [tab. 1.c]. Instead, the uniform sampling ADCs guarantee measurements of $\varepsilon_r$ on concretes with a low $\sigma$ (type D) [fig. 4.d]. In fact, it is sufficient to have $n_{min}=8$, even if the minimum value of frequency $f_{min}$ is too close to the end of the band $B$. In the best case, now a high speed ADC only similar to the *NI PCI(PXI)-5154*, with sampling rate $f_S=2GHz$, it results $f_{min}=94.953kHz$, such that $\Delta\sigma/\sigma(f_{min})=8.556\cdot 10^{-3}$ and $\Delta\varepsilon_r/\varepsilon_r(f_{min})=0.15$ [besides $f_{opt}=607.522kHz$, where $\Delta\sigma/\sigma(f_{opt})=0.033$, $\Delta\varepsilon r/\varepsilon_r(f_{opt})=0.017$]. Then, it is necessary to have $n_{min}=12$ to work at least in a small band $[f_{min},B]$. In the best case, again a high speed ADC similar to the *NI PCI(PXI)-5124*, it is $f_{min}=28.273kHz$, such that $\Delta\sigma/\sigma(f_{min})=1.059\cdot 10^{-3}$ [besides $f_{opt}=165.329kHz$, where $\Delta\sigma/\sigma(f_{opt})=4.346\cdot 10^{-3}$, $\Delta\varepsilon_r/\varepsilon_r(f_{opt})= 8.19\cdot 10^{-3}$] [tab. 1.d] [Settimi et al, 2009, b]. In particular, an high density uniform ADC only similar to the *NI PCI-6110*, with number of bits $n=12$ and sampling rate $f_S=5MHz$, does not allow to perform measurements till $B=100kHz$: concerning the concretes characterized by an high conductivity $\sigma$ (type C), only in the frequency band $[f_{min}=42.099kHz, f_{max}=95.531kHz]$, such that $\Delta\sigma/\sigma(f_{min})=0.034$ and $\Delta\sigma/\sigma(f_{max})=0.079$ [besides $f_{opt}=61.932kHz$, where $\Delta\sigma/\sigma(f_{opt})=0.051$, $\Delta\varepsilon_r/\varepsilon_r(f_{opt})=0.085$]; as regards the concretes with a low $\sigma$ (type D), only in the band $[f_{min}=35.085kHz, f_{max}=85.396kHz]$, such that $\Delta\sigma/\sigma(f_{min})=0.029$ and $\Delta\sigma/\sigma(f_{max})=0.071$ [besides $f_{opt}=53.271kHz$, where $\Delta\sigma/\sigma(f_{opt})=0.044$, $\Delta\varepsilon_r/\varepsilon_r(f_{opt})=0.082$].
Finally, a low cost ADC similar to the *NI USB-5133*, with $n=8$ and $f_S=100MHz$, guarantees measurements of $\rho$ and $\varepsilon_r$ which range from $(\rho_{min}=657.905\Omega\cdot m, \varepsilon_{r,max}=81)$ to $(\rho_{max}=20k\Omega\cdot m, \varepsilon_{r,min}=2.31)$, such that $\Delta\rho/\rho=0.013$, corresponding only to the unpolluted freshwater lakes with an electrical resistivity $\approx 1000\Omega\cdot m$ and a dielectric permittivity $\approx 80$ [Edwards, 1998]. Instead, a low cost ADC similar to the *NI PCI(PXI)-5105*, with $n=12$ and $f_S=60MHz$ [fig. 5.a], guarantees measurements of $\rho$ and $\varepsilon_r$ which range from $(\rho_{min}=160.044\Omega\cdot m, \varepsilon_{r,max}=81)$, such that $\Delta\rho/\rho=7.181\cdot 10^{-3}$, to $(\rho_{max}=20\ k\Omega\cdot m, \varepsilon_{r,min}=1)$, such that $\Delta\rho/\rho=7.397\cdot 10^{-3}$ and $\Delta\varepsilon_r/\varepsilon_r=0.034$, so including also the arid sand deserts without vegetation with a resistivity $>20000\Omega\cdot m$ and a permittivity $\approx 3$ [Edwards, 1998], and especially various types of concretes across the types C to D.

### 1.2. IQ sampling ADCs.

If the RESPER, of frequency band $B=100kHz$, is in galvanic contact with the subjacent medium, i.e. $h=0$, then, in order to measure the electrical conductivity $\sigma$ and the dielectric permittivity $\varepsilon_r$ with inaccuracies below a prefixed limit ($10\%$), the performances of an IQ sampling ADC are preferable when compared to an uniform sampling ADC. Under the same operating conditions, an uniform sampling can afford the advantage of the smallest resolution (i.e. $n=8$) to perform measurements on surfaces characterized by a low conductivity (i.e. the concretes), with the disadvantage to reduce the working band of frequency; in fact the band $B$ is higher than the maximum value of frequency $f_{max}$ which allows an inaccuracy $\Delta\varepsilon_r/\varepsilon_r$ in the measurement of $\varepsilon_r$ below the limit $10\%$. Instead, an IQ sampling can afford three advantages: first, over $B$, the inaccuracies $\Delta\sigma/\sigma$ and $\Delta\varepsilon_r/\varepsilon_r$, respectively for $\sigma$ and $\varepsilon_r$, are generally smaller by half an order of magnitude at least; second, the minimum value of frequency $f_{min}$ which allows an inaccuracy $\Delta\varepsilon_r/\varepsilon_r(f)$ below $10\%$ is generally slightly lower; third, the maximum value of frequency $f_{max}$ is always higher than $B$; as one and only disadvantage, using the IQ sampling, the optimal frequency $f_{opt}$ which minimizes $\Delta\varepsilon_r/\varepsilon_r(f)$ is generally higher by half a decade of LF-MF frequency, at least, compared to the uniform sampling [Settimi et al, 2009, b].

In fact, an IQ sampling guarantees to perform measurements on terrestrial soils characterized by an high conductivity $\sigma$ (type A) [fig. 4.a], assuming necessarily $n_{min}=18$ [tab. 1.a] as minimal number of bits, too high to make the measuring system insensitive to the external electrical noise [Settimi et al, 2009, b]; and on terrestrial soils of low $\sigma$ (type B) [fig. 4.b], with $n_{min}=12$, such that $f_{min}=94.228kHz$, where $\Delta\sigma/\sigma(f_{min})=9.819\cdot10^{-4}$, being $B=100kHz$, where $\Delta\sigma/\sigma(B)=9.825\cdot10^{-4}$ and $\Delta\varepsilon_r/\varepsilon_r(B)=0.089$ [besides $f_{opt}=1.459MHz$, where $\Delta\sigma/\sigma(f_{opt})=2.092\cdot10^{-3}$, $\Delta\varepsilon_r/\varepsilon_r(f_{opt})= 2.122\cdot10^{-3}$] [tab. 1.b] [Settimi et al, 2009, b]. Instead, the IQ sampling guarantees measurements on concretes with an high conductivity $\sigma$ (type C) [fig. 4.c], assuming $n_{min}=12$ as minimal number of bits, such that $f_{min}=33.292kHz$, where $\Delta\sigma/\sigma(f_{min})=9.813\cdot10^{-4}$, being $\Delta\sigma/\sigma(B)=1.017\cdot10^{-3}$ and $\Delta\varepsilon_r/\varepsilon_r(B)=0.012$ [besides $f_{opt}=547.144kHz$, where $\Delta\sigma/\sigma(f_{opt})=2.092\cdot10^{-3}$, $\Delta\varepsilon_r/\varepsilon_r(f_{opt})=1.886\cdot10^{-3}$][tab. 1.c]; and on concretes of low $\sigma$ (type D) [fig. 4.d], with $n_{min}=12$, so that $f_{min}=28.268kHz$, where $\Delta\sigma/\sigma(f_{min})=9.819\cdot10^{-4}$, being $\Delta\sigma/\sigma(B)=1.039\cdot10^{-3}$ and $\Delta\varepsilon_r/\varepsilon_r(B)=9.14\cdot10^{-3}$ [besides $f_{opt}=437.637kHz$, where $\Delta\sigma/\sigma(f_{opt})=2.091\cdot10^{-3}$, $\Delta\varepsilon_r/\varepsilon_r(f_{opt})= 2.121\cdot10^{-3}$][tab. 1.d][Settimi et al, 2009, b].

Moreover, just in a laboratory which has an anechoic chamber, shielded from the external electrical noise, the quadrupolar probe can be connected to an uniform or IQ sampling ADC, characterized by an high resolution, in order to perform measurements on samples of various terrestrial soils, drawn from the outside environment [Settimi et al, 2009, b]. In fact, for both the uniform and IQ sampling, it is necessary to have $n_{min}=20$, as minimal number of bits, to measure, with inaccuracies below the limit of $10\%$, the electrical resistivity $\rho$ and the dielectric permittivity $\varepsilon_r$ which range from $(\rho_{min}=21.941\Omega\cdot m,\ \varepsilon_{r,max}=81)$ to $(\rho_{max}=20k\Omega\cdot m,\ \varepsilon_{r,min}=1)$, corresponding to various terrestrial soils included in the following list [Edwards, 1998]: pastoral, low hills, fertile soil $(\rho\approx80\Omega\cdot m,\ \varepsilon_r\approx15)$; flat, marshy, densely wooded in places $(\approx130\Omega\cdot m,\ \approx13)$; pastoral, heavy clay soils, hills $(\approx250\Omega\cdot m,\ \approx12)$; pastoral, medium hills with forestation $(\approx270\Omega\cdot m,\ \approx12)$; rocky, sandy with some rainfall or some vegetation $(\approx500\Omega\cdot m,\ \approx8)$; rocky soil, steep forested hills, streams $(\approx500\Omega\cdot m,\ \approx10)$; low-rise city suburbs, built-up areas, parks $(\approx1000\Omega\cdot m,\ \approx6)$; high-rise city centres, industrial areas $(\approx3000\Omega\cdot m,\ \approx4)$. Then, it is necessary to have $n_{min}=24$ to measure, with inaccuracies below $10\%$, the resistivity $\rho$ and the permittivity $\varepsilon_r$ typical of agricultural plains, streams and rich loam soil $(\rho\approx30\Omega\cdot m,\ \varepsilon_r\approx20)$. Finally, the above list does not include the sea water, away from river estuaries $(\approx0.22\Omega\cdot m,\ \approx81)$ [Edwards, 1998]. Instead, an IQ sampling, with number of bits $n=8$, guarantees measurements of $\rho$ and $\varepsilon_r$ ranging from $(\rho_{min}=674.098\Omega\cdot m,\ \varepsilon_{r,max}=81)$ to $(\rho_{max}=20k\Omega\cdot m,\ \varepsilon_{r,min}=2.41)$, where $\Delta\rho/\rho=0.017$, so, compared to the uniform sampling, a less wide variety of unpolluted freshwater lakes with a major inaccuracy in the measurements of the electrical resistivity and dielectric permittivity. Then, an IQ sampling, with $n=12$ [fig. 5.b], guarantees measurements of $\rho$ and $\varepsilon_r$ ranging from $(\rho_{min}=154.937\Omega\cdot m,\ \varepsilon_{r,max}=81)$, where $\Delta\rho/\rho=9.809\cdot10^{-4}$, to $(\rho_{max}=20\ k\Omega\cdot m,\ \varepsilon_{r,min}=1)$, where $\Delta\rho/\rho=1.017\cdot10^{-3}$ and $\Delta\varepsilon_r/\varepsilon_r=0.022$, so, compared to the uniform sampling, even more arid sand deserts without vegetation with a minor inaccuracy for the resistivity and permittivity, and especially more types of concretes across the types C and D.

Finally, when the quadrupole is in capacitive contact with the medium, being designed to perform optimal measurements, if an IQ sampling is used, then, almost independently from the probe configuration [Settimi et al, 2009, a], just the high resolution $n_{min}=18$ allows to measure the dielectric permittivity $\varepsilon_r$, with inaccuracy $\Delta\varepsilon_r/\varepsilon_r$, of terrestrial soils characterized by an high electrical conductivity $\sigma$ (type A), with inaccuracy $\Delta\sigma/\sigma$ [fig. 4.a.bis]. The quadruple using the IQ sampling with $n_{min}=18$ could work in both MF at most one decade lower or better if within the band $[f_{up}=1.024MHz,\ z_A=3.878MHz]$ and HF even three decades higher or better if within $[f_{low}=10.663MHz,\ p_A=15.514MHz]$ [such that $\Delta\sigma/\sigma$ shows its minimum point in the upper limit $f_{up}$, so $\Delta\sigma/\sigma(f_{up})=2.527\cdot10^{-5}$ and $\Delta\varepsilon_r/\varepsilon_r(f_{up})=3.028\cdot10^{-4}$, while $\Delta\sigma/\sigma$ and $\Delta\varepsilon_r/\varepsilon_r$ are equal in the lower limit $f_{low}$, i.e. $\Delta\sigma/\sigma(f_{low})=\Delta\varepsilon_r/\varepsilon_r(f_{low})=3.194\cdot10^{-5}$][tab. 1.a.bis]. Instead, the minimal number of bits $n_{min}=12$ allows to measure the permittivity $\varepsilon_r$ of: terrestrial soils with a low conductivity $\sigma$ (type B) [fig. 4.b.bis], for both MF at most half a decade lower than the band $[f_{up}=124.768kHz,\ z_B=470.587kHz]$ and HF one decade higher than $[f_{low}=1.492MHz,\ p_B=1.882MHz]$ [such that $\Delta\sigma/\sigma(f_{up})=1.64\cdot10^{-3}$, $\Delta\varepsilon_r/\varepsilon_r(f_{up})=0.022$ and $\Delta\sigma/\sigma(f_{low})=\Delta\varepsilon_r/\varepsilon_r(f_{low})=2.311\cdot10^{-3}$] [tab. 1.b.bis]; concretes with an high $\sigma$ (type C) [fig. 4.c.bis], for both (MF at most about half a decade lower than) the band $[f_{up}=46.648kHz,\ z_C=176.47kHz]$ and (HF about one decade higher than) $[f_{low}=499.469kHz,\ p_C=705.881kHz]$ [such that $\Delta\sigma/\sigma(f_{up})=1.622\cdot10^{-3}$, $\Delta\varepsilon_r/\varepsilon_r(f_{up})=0.02$ and

$\Delta\sigma/\sigma(f_{low})=\Delta\varepsilon_r/\varepsilon_r(f_{low})=2.092\cdot 10^{-3}$] [tab. 1.c.bis]; concretes with a low $\sigma$ (type D) [fig. 4.d.bis], for both the band $[f_{up}=37.43kHz,\ z_D=141.176kHz]$ and $[f_{low}=477.58kHz,\ p_D=564.705kHz]$ [such that $\Delta\sigma/\sigma(f_{up})=1.64\cdot 10^{-3}$, $\Delta\varepsilon_r/\varepsilon_r(f_{up})=0.022$ and $\Delta\sigma/\sigma(f_{up})=\Delta\varepsilon_r/\varepsilon_r(f_{low})=2.311\cdot 10^{-3}$] [tab. 1.d.bis].

### 1.3. Lock-in amplifiers.

If the RESPER, of frequency band $B=100kHz$, is in galvanic contact with the subjacent medium, then, in order to measure the electrical conductivity $\sigma$ and the dielectric permittivity $\varepsilon_r$ with inaccuracies below a prefixed limit (*10%*), the performances of a lock-in amplifier only similar to the *SR810* (and *SR830*) are preferable when compared to a IQ sampling ADC with number of bits $n_{IQ}=12$. In fact, the lock-in amplifier is specified by a voltage sensitivity $\Delta V/V=2\cdot 10^{-9}$ and a degree phase resolution $\Delta\varphi=0.01^\circ$, so it guarantees inaccuracies for the transfer impedance in modulus, i.e. $\Delta|Z|/|Z|_{lock\text{-}in}=2\cdot\Delta V/V=4\cdot 10^{-9}$, and in phase, i.e. $\Delta\Phi/\Phi_{lock\text{-}in}=\Delta\varphi/360^o=2.778\cdot 10^{-5}$, that are smaller than the inaccuracies due to IQ sampling ADC characterized by a number of bits $n_{IQ}=12$, respectively of five and one order of magnitude, i.e. $\Delta|Z|/|Z|_{IQ}(n_{IQ})=\Delta\Phi/\Phi_{IQ,max}(n_{IQ})=1/2^{nIQ}\approx 2.441\cdot 10^{-4}$ [Settimi et al, 2009, b]. Connecting the lock-in amplifier, the inaccuracy $\Delta\varepsilon_r/\varepsilon_r(f)$ in the measurement of $\varepsilon_r$, as function of the frequency $f$, is like a concave upward parabola which, compared to the IQ sampling ADC, shows a squashed shape around its minimum, being left shifted towards LF and shifted downwards until smaller values. So, the lock-in amplifier, more than the IQ sampling, guarantees: first, an optimal frequency $f_{opt}$, minimizing the inaccuracy $\Delta\varepsilon_r/\varepsilon_r(f)$ for $\varepsilon_r$, which is lower of one MF decade, generally falling on middle frequencies (*100kHz-1MHz*) and, for almost all the concretes, within the band $B=100kHz$; second, a minimal frequency $f_{min}$, allowing an inaccuracy $\Delta\varepsilon_r/\varepsilon_r(f)$ below the limit of *10%*, which is smaller even of two LF decades, generally falling on low frequencies ($\approx 1kHz$) and, for the concretes, on even lower frequencies ($\approx 100Hz$). Then, the inaccuracy $\Delta\varepsilon_r/\varepsilon_r(f)$ is generally smaller by one magnitude order at least and three orders at most, while the inaccuracy $\Delta\sigma/\sigma(f)$ in the measurement of $\sigma$ can be approximated to a constant, i.e. $\Delta\sigma/\sigma\approx 2\cdot\Delta\Phi/\Phi_{lock\text{-}in}=5.556\cdot 10^{-5}$, since the inaccuracy for $|Z|$, i.e. $\Delta|Z|/|Z|_{lock\text{-}in}=2\cdot\ \Delta V/V=4\cdot 10^{-9}$, is too little to contribute to the inaccuracy $\Delta\sigma/\sigma(f)$ of $\sigma$, so $\Delta\sigma/\sigma(f)=[1/S_{|Z|,\sigma}(f)]\cdot\Delta|Z|/|Z|_{lock\text{-}in}+\ [1/S_{\Phi,\sigma}(f)]\cdot\Delta\Phi/\Phi_{lock\text{-}in}\approx[1/S_{\Phi,\sigma}(f)]\cdot\Delta\Phi/\Phi_{lock\text{-}in}$, and the sensitivity function of $\Phi$ relative to $\sigma$, i.e. $S_{\Phi,\sigma}(f)\approx\langle S_{\Phi,\sigma}(f)\rangle=1/2$, is almost constant with the frequency $f$ assuming the mean value $1/2$, so $\Delta\sigma/\sigma(f)\approx 2\cdot\ \Delta\Phi/\Phi_{lock\text{-}in}$ [Settimi et al, 2009, a].
Differently from the IQ sampling ADC, with number of bits $n_{IQ}=12$, the lock-in amplifier guarantees to perform measurements on terrestrial soils characterized by an high conductivity $\sigma$ (type A) [fig. 4.a], so that $f_{min}=2.9kHz$, where $\Delta\sigma/\sigma(f_{min})=5.556\cdot 10^{-5}$, being $B=100kHz$, where $\Delta\sigma/\sigma(B)=5.557\cdot 10^{-5}$ and $\Delta\varepsilon_r/\varepsilon_r(B)=1.439\cdot 10^{-4}$ [besides $f_{opt}=1.199MHz$, where $\Delta\sigma/\sigma(f_{opt})=5.611\cdot 10^{-5}$, $\Delta\varepsilon_r/\varepsilon_r(f_{opt})=\ 6.101\cdot 10^{-5}$] [tab. 1.a]; and on terrestrial soils of low $\sigma$ (type B) [fig. 4.b], so that $f_{min}=379.08Hz$, where $\Delta\sigma/\sigma(f_{min})=5.556\cdot 10^{-5}$, being $\Delta\sigma/\sigma(B)=5.582\cdot 10^{-5}$ and $\Delta\varepsilon_r/\varepsilon_r(B)=7.121\cdot 10^{-5}$ [besides $f_{opt}=145.489kHz$, where $\Delta\sigma/\sigma(f_{opt})=5.611\cdot 10^{-5}$, $\Delta\varepsilon_r/\varepsilon_r(f_{opt})=7.081\cdot 10^{-5}$] [tab. 1.b]. In comparison to IQ sampling ADC, with $n_{IQ}=12$, the lock-in amplifier guarantees more accurate measurements on concretes with an high conductivity $\sigma$ (type C) [fig. 4.c], so that $f_{min}=134.02Hz$, where $\Delta\sigma/\sigma(f_{min})=5.556\cdot 10^{-5}$, being $B=100kHz$, where $\Delta\sigma/\sigma(B)=\ 5.738\cdot 10^{-5}$ and $\Delta\varepsilon_r/\varepsilon_r(B)=6.393\cdot 10^{-5}$ [besides $f_{opt}=54.558kHz$, where $\Delta\sigma/\sigma(f_{opt})=5.611\cdot 10^{-5}$, $\Delta\varepsilon_r/\varepsilon_r(f_{opt})=6.295\cdot 10^{-5}$] [tab. 1.c]; and on concretes of low $\sigma$ (type D) [fig. 4.d], so that $f_{min}=113.724Hz$, where $\Delta\sigma/\sigma(f_{min})=5.556\cdot 10^{-5}$, being $\Delta\sigma/\sigma(B)=5.839\cdot 10^{-5}$ and $\Delta\varepsilon_r/\varepsilon_r(B)=7.311\cdot 10^{-5}$ [besides $f_{opt}=43.647kHz$, where $\Delta\sigma/\sigma(f_{opt})=\ 5.611\cdot 10^{-5}$, $\Delta\varepsilon_r/\varepsilon_r(f_{opt})=7.081\cdot 10^{-5}$] [tab. 1.d].
Moreover, if the quadrupolar probe is connected to a lock-in amplifier with specifications similar to the *SR810*'s ones, then the measurements on various terrestrial soils can be performed extemporaneously and "in situ", differently from connecting an IQ sampling ADC, characterized by an high bit resolution, when the samples of those terrestrial soils must be drawn from the outside environment to be analyzed in a laboratory which has an anechoic chamber, shielded from the external electrical noise [Settimi et al, 2009, b]. The lock-in amplifier can not yet measure the resistivity $\rho$ and the permittivity $\varepsilon_r$ of the sea water, away from river estuaries [Edwards, 1998], with inaccuracies $\Delta\rho/\rho$ and $\Delta\varepsilon_r/\varepsilon_r$ below *10%*, but, in comparison to the IQ sampling with a number of bit $n=24$, it can perform optimal measurements on even more conductive agricultural plains, streams and richer loam soil [Edwards, 1998], though with inaccuracies generally higher of two magnitude orders. In fact, the lock-in amplifier guarantees to measure $\rho$ and $\varepsilon_r$ ranging from $(\rho_{min}=21.941\Omega\cdot m,\ \varepsilon_{r,max}=81)$, such that $\Delta\rho/\rho=5.557\cdot 10^{-5}$ and $\Delta\varepsilon_r/\varepsilon_r=1.371\cdot 10^{-4}$, to $(\rho_{max}=20k\Omega\cdot m,\ \varepsilon_{r,min}=1)$, such that $\Delta\rho/\rho=5.738\cdot 10^{-5}$ and $\Delta\varepsilon_r/\varepsilon_r=1.151\cdot 10^{-4}$ [where $(\rho_{opt}=266.138\Omega\cdot m,\ \varepsilon_{r,opt}=81)$ such that $\Delta\rho/\rho=5.611\cdot 10^{-5}$ and $\Delta\varepsilon_r/\varepsilon_r=5.735\cdot 10^{-5}$] [fig. 5.c]. Instead, an IQ sampling, with $n=24$, guarantees to measure $\rho$ and $\varepsilon_r$ in the same range from $(\rho_{min},\ \varepsilon_{r,max})$, such that $\Delta\rho/\rho=2.385\cdot 10^{-7}$ and $\Delta\varepsilon_r/\varepsilon_r=1.206\cdot 10^{-3}$, to $(\rho_{max},\ \varepsilon_{r,min})$, such that $\Delta\rho/\rho=2.483\cdot10^{-7}$ and $\Delta\varepsilon_r/\varepsilon_r=5.302\cdot10^{-6}$ [but where ($\rho_{opt}=2.669k\Omega\cdot m,\ \varepsilon_{r,opt}=81$) such that $\Delta\rho/\rho=5.108\cdot10^{-7}$ and $\Delta\varepsilon_r/\varepsilon_r=4.195\cdot10^{-7}$].

Finally, if the quadrupole is in capacitive contact with the medium, being designed to perform optimal measurements, then, almost independently from the probe configuration [Settimi et al, 2009, a], the performances due to the lock-in amplifier are generally preferable when compared to an IQ sampling ADC with high resolution [Settimi et al, 2009, b]. The upper limit $f_{up}$ for the MF is only slightly left shifted compared to an IQ sampling [being the upper limit $f_{up}$ defined as the frequency where the inaccuracy $\Delta\sigma/\sigma(f)$ in the measurement of the electrical conductivity $\sigma$ shows a point of minimum, i.e. $\partial_f\Delta\sigma/\sigma(f)|_{f=fup}=0$]. The lower limit $f_{low}$ for the HF is right (types A-C) or left (type B-D) shifted compared to an IQ sampling [being the lower limit $f_{low}$ defined as the frequency where the inaccuracy $\Delta\sigma/\sigma(f)$ for the conductivity $\sigma$ is equal to the inaccuracy $\Delta\varepsilon_r/\varepsilon_r(f)$ for $\varepsilon_r$, i.e. $\Delta\sigma/\sigma(f_{low})=\Delta\varepsilon_r/\varepsilon_r(f_{low})$]. Numerically: type A, $[f_{up}=825.975kHz,\ f_{low}=22.402MHz]$ [tab. 1.a.bis]; type B, $[f_{up}=100.508kHz,\ f_{low}=780.277kHz]$ [tab. 1.b.bis]; type C, $[f_{up}=37.606kHz,\ f_{low}=546.487kHz]$ [tab. 1.c.bis]; type D, $[f_{up}=30.152kHz,\ f_{low}=234.083kHz]$ [tab. 1.d.bis]. Unfortunately, in the case of the terrestrial soils characterized by an high electrical conductivity $\sigma$ (type A), the inaccuracy $\Delta\varepsilon_r/\varepsilon_r(f)$, as function of the frequency $f$, is shifted upwards to values larger of almost one magnitude order, due to an highest $\sigma$, so that the inaccuracy values increase of almost one order from $10^{-5}$ until $10^{-4}$. The band of the MF $f<f_{up}$ maintains almost invariant, a range of about one decade, as the HF band $f>f_{low}$, a range of about three decades, compared to the IQ sampling with number of bits $n=18$. Numerically: type A, $\Delta\sigma/\sigma(f_{up})=1.025\cdot10^{-4}$, $\Delta\varepsilon_r/\varepsilon_r(f_{up})=8.192\cdot10^{-4}$ and $\Delta\sigma/\sigma(f_{low})=\Delta\varepsilon_r/\varepsilon_r(f_{low})=1.925\cdot10^{-4}$ [fig. 4.a.bis]. Rather, in the cases of terrestrial soils with a low conductivity $\sigma$ (type B), concretes of high (type C) and low (type D) $\sigma$, the inaccuracy $\Delta\varepsilon_r/\varepsilon_r(f)$, as function of the frequency $f$, is shifted downwards to values just higher of one magnitude order and slightly towards lower frequencies, due to the effect of the lock-in amplifier, so that the inaccuracy values decrease at least of one order from $10^{-3}$ until $10^{-4}$. Even if the band of the MF $f<f_{up}$ is slightly broadened, anyway the HF band $f>f_{low}$ is broadened until two-three decades at most (type B-D) or one decade at least (type C), compared to the IQ sampling of $n=12$. Numerically: type B, $\Delta\sigma/\sigma(f_{up})=1.041\cdot10^{-4}$, $\Delta\varepsilon_r/\varepsilon_r(f_{up})=9.487\cdot10^{-4}$ and $\Delta\sigma/\sigma(f_{low})=\Delta\varepsilon_r/\varepsilon_r(f_{low})=1.037\cdot10^{-4}$ [fig. 4.b.bis]; type C, $\Delta\sigma/\sigma(f_{up})=1.029\cdot10^{-4}$, $\Delta\varepsilon_r/\varepsilon_r(f_{up})=8.466\cdot10^{-4}$ and $\Delta\sigma/\sigma(f_{low})=\Delta\varepsilon_r/\varepsilon_r(f_{low})=1.183\cdot10^{-4}$ [fig. 4.c.bis]; type D, $\Delta\sigma/\sigma(f_{up})=1.041\cdot10^{-4}$, $\Delta\varepsilon_r/\varepsilon_r(f_{up})=9.478\cdot10^{-4}$ and $\Delta\sigma/\sigma(f_{low})=\Delta\varepsilon_r/\varepsilon_r(f_{low})=1.037\cdot10^{-4}$ [fig. 4.d.bis].

## 2. Characteristic geometrical dimensions of the RESPER probe.

In ref. [Settimi et al, 2009, b], we have demonstrated, once fixed the input resistance $R_{in}$ of the amplifier stage and selected the working frequency $f$ of the RESPER [fig. 1], which falls in a LF band starting from the minimum value of frequency $f_{min}$ for operating conditions of galvanic contact or in a MF-HF band from the lower limit $f_{low}$ for capacitive contact, that the minimal radius $r(R_{in},f_{min})$ of the electrodes can be designed, as it depends only on the resistance $R_{in}$ and the frequency $f$, being a function inversely proportional to both $R_{in}$ and $f$,

$$r \simeq \frac{1}{(2\pi)^2 \varepsilon_0 R_{in} f} \quad , \; f \geq f_{\min}, f_{low} . \tag{2.1}$$

Moreover, known the minimal number of bits $n_{min}$ for the uniform or IQ sampling ADC [fig. 2], which allows an inaccuracy in the measurement of the dielectric permittivity $\varepsilon_r$ below the prefixed limit *10%*, also the electrode-electrode distance $L(r,n_{min})$ can be projected, as it depends only on the radius $r(R_{in},f_{min})$ and the number of bits $n_{min}$, being directly proportional to $r(R_{in},f_{min})$ and increasing as the exponential function $2^{nmin}$ of $n_{min}$,

$$L \propto r \cdot 2^{n_{\min}} . \tag{2.2}$$

Finally, the radius $r(R_{in},f_{min})$ maintains invariant whether the quadrupole probe assumes the square [fig. 3.a] or the linear (Wenner's) [fig. 3.b] configuration, while, known the number of bits $n_{min}$ of the ADC, then the distance $L_S(r,n_{min})$ in the square configuration must be smaller of a factor ($2-2^{1/2}$) compared to the corresponding one $L_W(r,n_{min})$ in the Wenner's case,

$$L_S = (2-\sqrt{2}) \cdot L_W = (2-\sqrt{2}) \cdot r \cdot 2^{n_{\min}} . \tag{2.3}$$

We suppose that the quadrupole is connected to an IQ sampling ADC, having engineered an electronic filter to remove the electrical noise of the external environment. In order to maintain the input resistance $R_{in}$ of the amplifier stage around easily available values with magnitude order of *1-250MΩ* so below the almost impracticable limit of *1GΩ*, the square configuration is preferable when compared to the linear (Wenner's) case [Settimi et al, 2009, a].

In fact, the square configuration, a part from the terrestrial ground characterized by an high electrical conductivity $\sigma$ (type A), is fit for analyzing all the other surfaces, as the terrestrial soils with a low conductivity $\sigma$ (type B) and the concretes of high (type C) or low (type D) $\sigma$, both in galvanic and in capacitive contact, though generally one should take the greatest care in designing the electrodes of this configuration, which require a non negligible radius around *417μm*, with the intent to reduce their interactions of capacitive coupling. Assume that the probe looks like a small electrical carpet designed to perform an electrical spectrography on road paving and buildings, i.e. a bi-dimensional array of characteristic geometrical dimension about $L \approx 1m$ both for terrestrial soils and concretes. With surfaces of type A, the probe is fit presenting only a capacitive contact, so that the amplifier stage should be projected with an input resistance $R_{in}=41.19M\Omega$, while the electrodes with a radius $r \approx 6.514\mu m$, so, only in this case, without any interaction between the electrodes. Type B [fig. 6.a], the probe is fit presenting both a galvanic and a capacitive contact, so, in the first case, the amplifier must be projected with a resistance $R_{in}=72.84M\Omega$, and the electrodes with a radius $r \approx 416.813\mu m$ [Settimi et al, 2009, b], while, in the second case, respectively $R_{in}=4.6M\Omega$ and $r \approx 416.836\mu m$ [tab. 2.a]. Type C [fig. 6.b], in galvanic contact, the amplifier must be projected with a resistance $R_{in}=206.1M\Omega$, and the electrodes with a radius $r \approx 416.94\mu m$, while, in capacitive contact, $R_{in}=13.74M\Omega$ and $r \approx 416.866\mu m$ respectively [tab. 2.b]. Type D [fig. 6.c], in galvanic contact, $R_{in}=242.8M\Omega$ and $r \approx 416.819\mu m$ [Settimi et al, 2009, b], while, in capacitive contact, $R_{in}=14.37M\Omega$ and $r \approx 416.858\mu m$ respectively [tab. 2.c].

Instead, the Wenner's configuration, a part from the terrestrial ground characterized by a low electrical conductivity $\sigma$ (type B), is fit for analyzing all the other surfaces, as the terrestrial soils with an high conductivity $\sigma$ (type A) and the concretes of high (type C) or low (type D) $\sigma$, only in capacitive contact, with the advantage that generally the electrodes of this other configuration require a radius around *10-100μm*, small enough to avoid the occurrence of their capacitive coupling interactions. Assume that the probe is a coaxial cable designed for geophysical prospecting, i.e. a one-dimensional array of length $L_{tot}=3L$ about $\approx 10m$, so only for terrestrial soils. With surfaces of type A, the probe is fit presenting only a capacitive contact, so that the amplifier stage should be projected with an input resistance $R_{in}=21.09M\Omega$, while the electrodes with a radius $r \approx 12.721\mu m$. Type B [fig. 6.d], the probe is fit presenting both a galvanic and a capacitive contact, so, in the first case, the amplifier must be projected with a resistance $R_{in}=37.3M\Omega$, and the electrodes with a radius $r \approx 813.959 \mu m$ [Settimi et al, 2009, b], while, in the second case, $R_{in}=2.356M\Omega$ and $r=813.856\mu m$ respectively, so, compared to the previous cases, with an even stronger interaction between the electrodes [tab. 2.d]. Finally, assume that the probe is a coaxial cable designed to perform an electrical spectrography on buildings, i.e. a one-dimensional array of length about $L_{tot} \approx 1m$, now only for the concretes. Type C, only in capacitive contact, the amplifier must be projected with a resistance $R_{in}=70.38M\Omega$, and the electrodes with a radius $r \approx 81.383\mu m$. Type D, only in capacitive contact, $R_{in}=73.6M\Omega$ and $r \approx 81.389\mu m$ respectively.

## 3. Conclusions.

The present paper has proposed a discussion of theoretical design and moved towards the development and engineering of a induction probe for electrical spectroscopy which acquires complex impedance in the field, filling the current technological gap.

Applying the same principle, but limited to the acquisition only of resistivity, there are various commercial instruments used in geology for investigating the first 2-5 meters underground both for the exploration of environmental areas and archaeological investigation [Samouëlian, 2005].

As regards the direct determination of the dielectric permittivity in subsoil, omitting geo-radar which provides an estimate by complex measurement procedures on radar-gram processing [Declerk, 1995][Sbartaï et al., 2006], the only technical instrument currently used is the so-called time-domain reflectometer (TDR), which utilizes two electrodes inserted deep in the ground in order to acquire this parameter for further analysis [Mojid et al., 2003][Mojid and Cho, 2004].

Since, in our previous papers, we had already addressed and resolved the conceptual problem for designing the "heart" of instrument, in this paper, we have completed the technical project [fig. 7] [tab. 3], in forecast of the two next aims.

First aim: the implementation of an hardware which can handle many electrodes, arranged such as to provide data which are related to various depths of investigation for a single pass of measurements; so, this hardware must be able to switch automatically the transmitting and measurement couples.

The RESPER could be connected to a bi-dimensional array formed by a grid of coplanar electrodes, where a multiplexer selects two transmitters and two receivers (geometrically similar to a small electrical *carpet* and employable by a single operator with both hands). Through a series of *4x16* multiplexers, connected to the *4* inputs of probe, one selects different contacts of the electrical *carpet* in automatic way through an USB port. In this way, shifting the position of measuring electrodes on the small *carpe*t, there is possibility to obtain a tomography without moving electrodes but simply by supporting the *carpet* and making to perform the operations of selection and measurement by a program of control and processing.

Otherwise, the probe could be connected to an uni-dimensional array consisting of a simple coaxial cable with antennas in line, where a muliplexer selects two transmitters and two receivers [only geometrically similar to *Geometrics Ohmmapper* [Walker and Houser, 2002] and dragged on soils or by a single operator or by a ground vehicle].

Second aim: the implementation of acquisition configurations, by an appropriate choice of the transmission frequency, for the different applications in which this instrument can be profitably used.

If the RESPER is connected to a DAS as an uniform or rather an IQ sampling ADC, then it could be commercialized for companies of building and road paving (ANAS), being employable for analyzing "in situ" only concretes.

The probe, connected to a small electrical *carpet*, performs a scan, adhering on large areas, as cements with high dielectric permittivity and draining asphalts. Then, the electrical *carpet*, being characterized by an area of $1m^2$, tests only the most superficial part, usually up to about *1m*, in particular for measurements of concretes. For manufactures of modern architecture and civil engineering, even if the small *carpet* investigates properties which are not related directly with the mechanical resistance of concrete, anyway they are related to its weaving, and therefore to its quality and actual state "in opera". Moreover, the *carpet* allows the quality check of asphalting, especially when porous asphalts are placed to avoid the effect of aquaplaning. This response lies mainly in dielectric permittivity, well known and constant for the mixture of asphalt, but depending on its content in rubber and air.

If the RESPER is connected to a DAS as a lock-in amplifier, then it could be commercialized for companies of geophysical prospecting, being employable for analyzing "in situ" even the terrestrial ground.

The probe, connected to a new instrument which could be named as *Ohm-Farad-Mapper* (*OFM*), is lift off ground, leading an electrical current through a system of antennas isolated from the terrestrial soil. Then, the probe analyzes large stretches of land, using an *OFM*, of length about *10m*, with antennas in line which provide a map of the soil, generally valid until the depth just of *10 m*. Moreover, the probe provides, as regards pedological field, data on both the lithology of surface lands and on their water content at the time of measurements, and, especially in view of that technological progress which names as precision agriculture, data on the composition of both the land and fertilizers.

Finally, but certainly not last in terms of importance, the timing i.e. daily productivity in terms of covered areas. The daily productivity in data acquisition is to be considered comparable with geo-radar, which, as known, acquires the radar-gram while it is dragged, normally with a speed of one walking pace (dozens of *cm/s* on manufactures, *1m* of profile per second on the surface of ground).

## References.

**Figures and captions.**

Figure 1

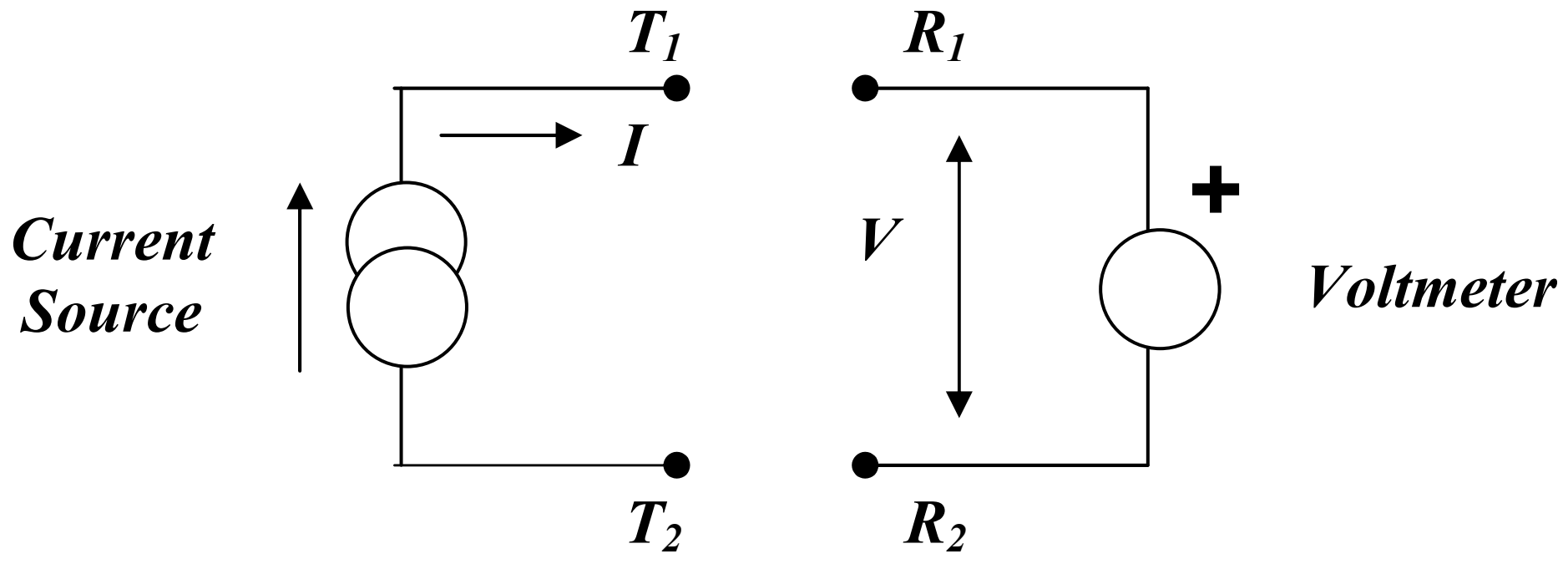

Figure 2

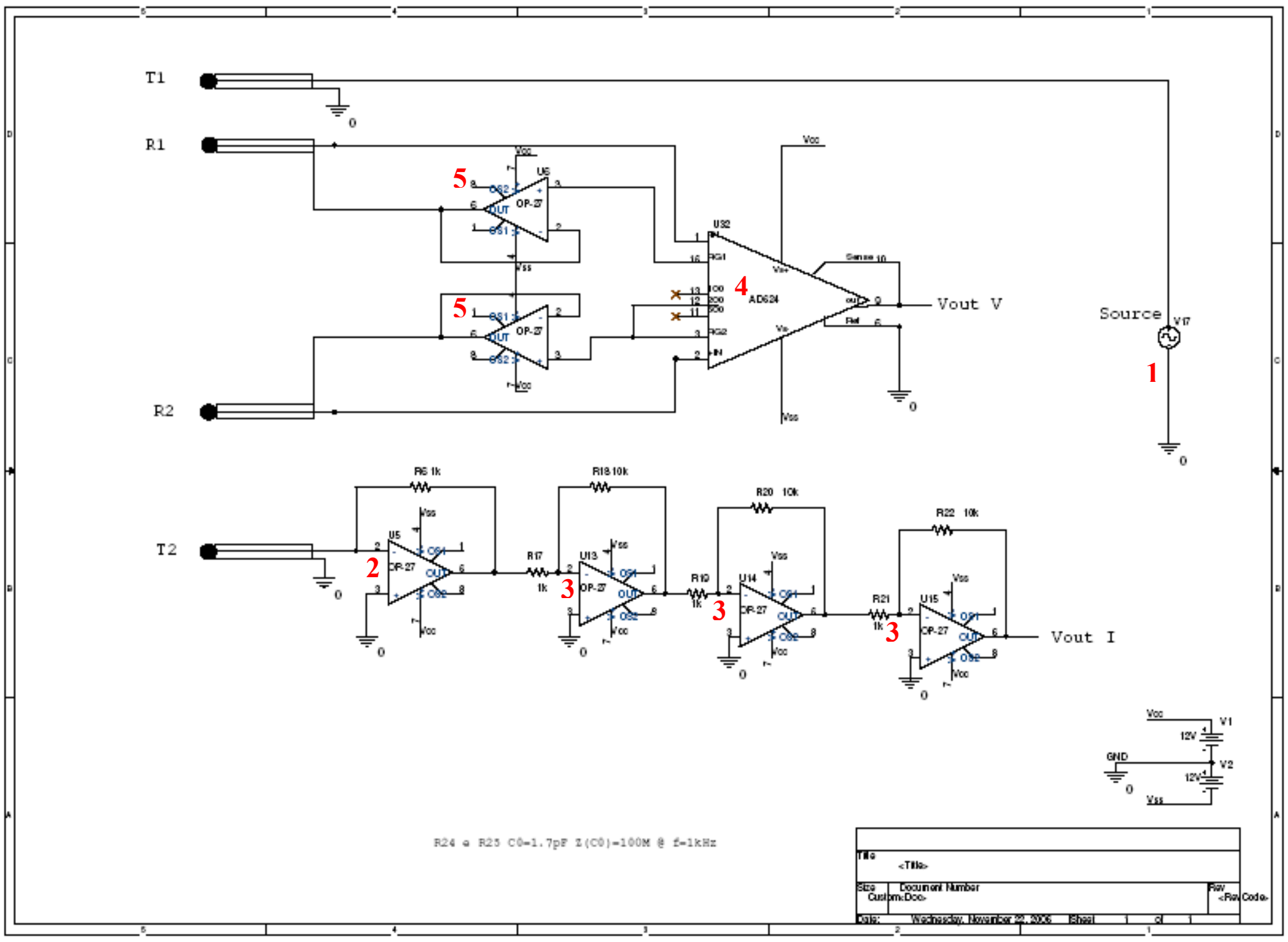

Figure 3.a

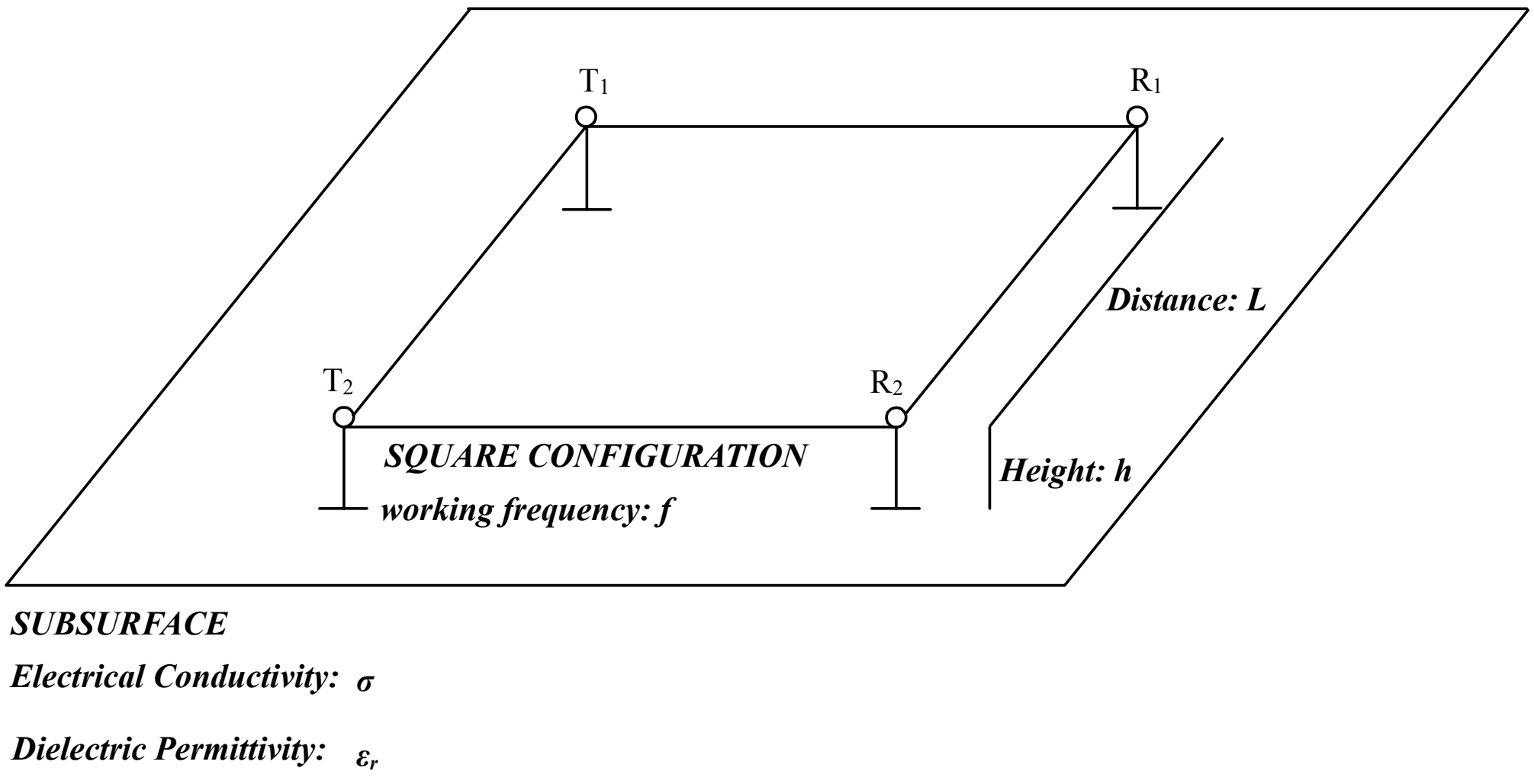

Figure 3.b

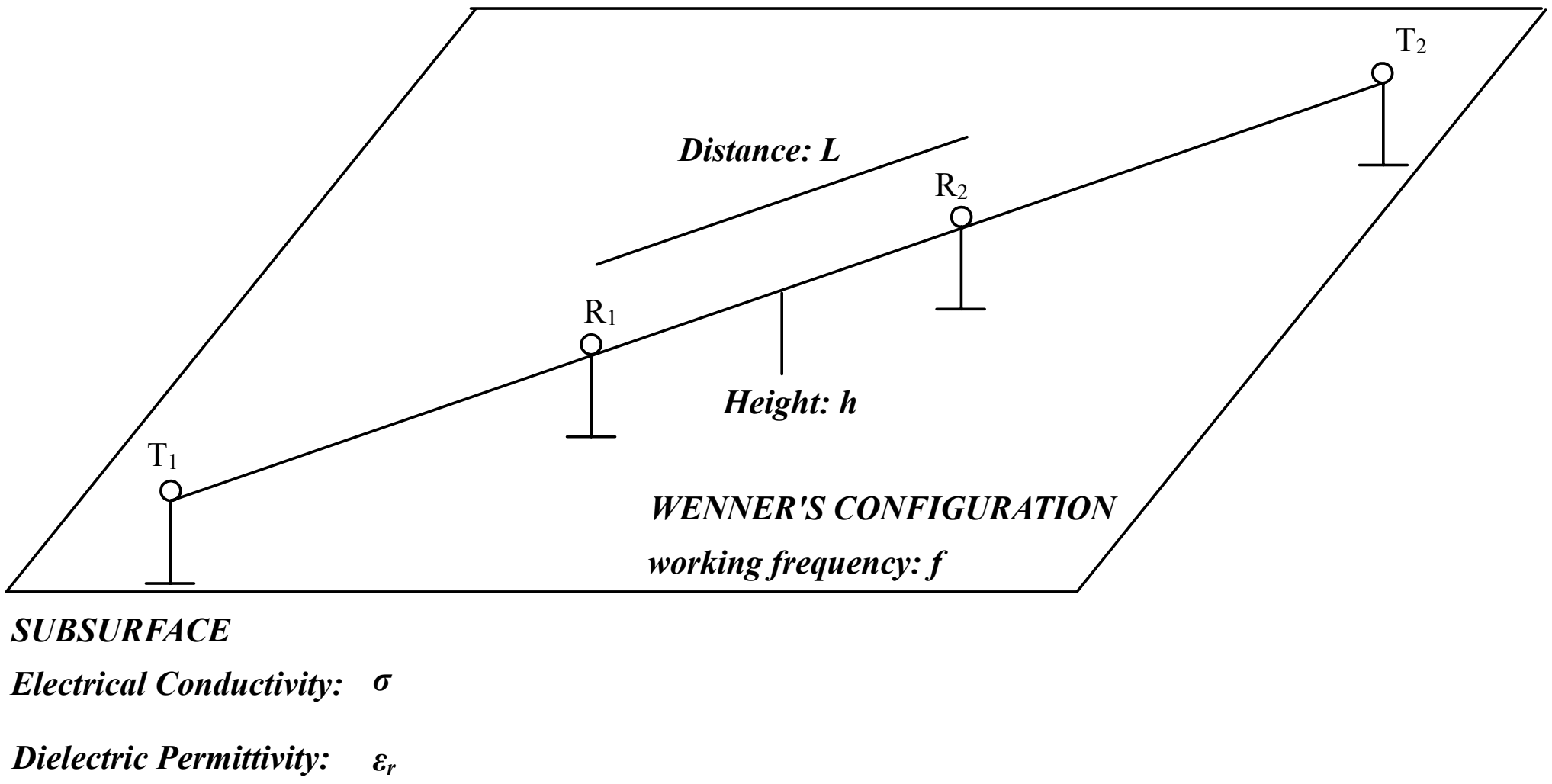

Figure. 4.a

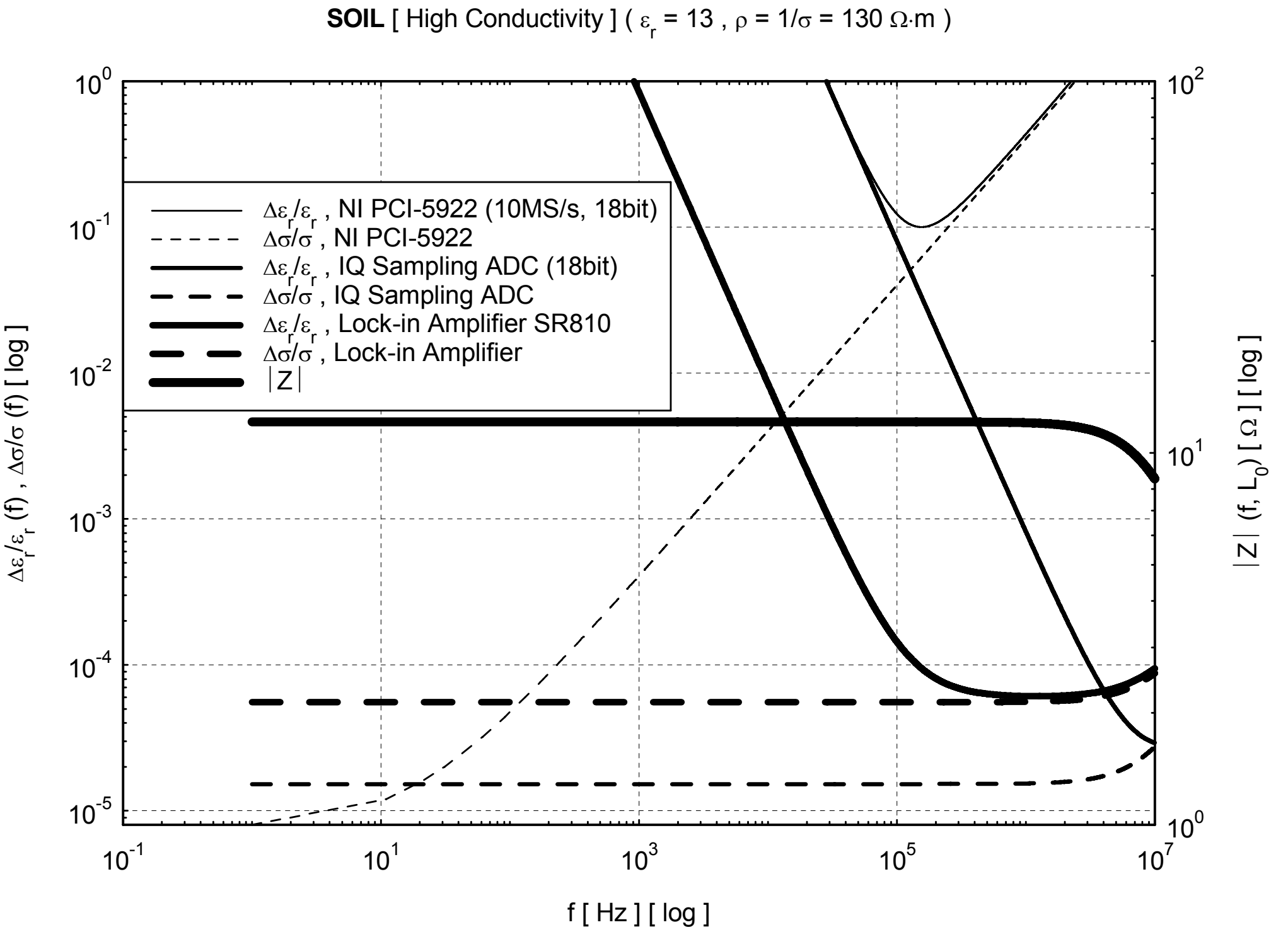

Figure 4.a.bis

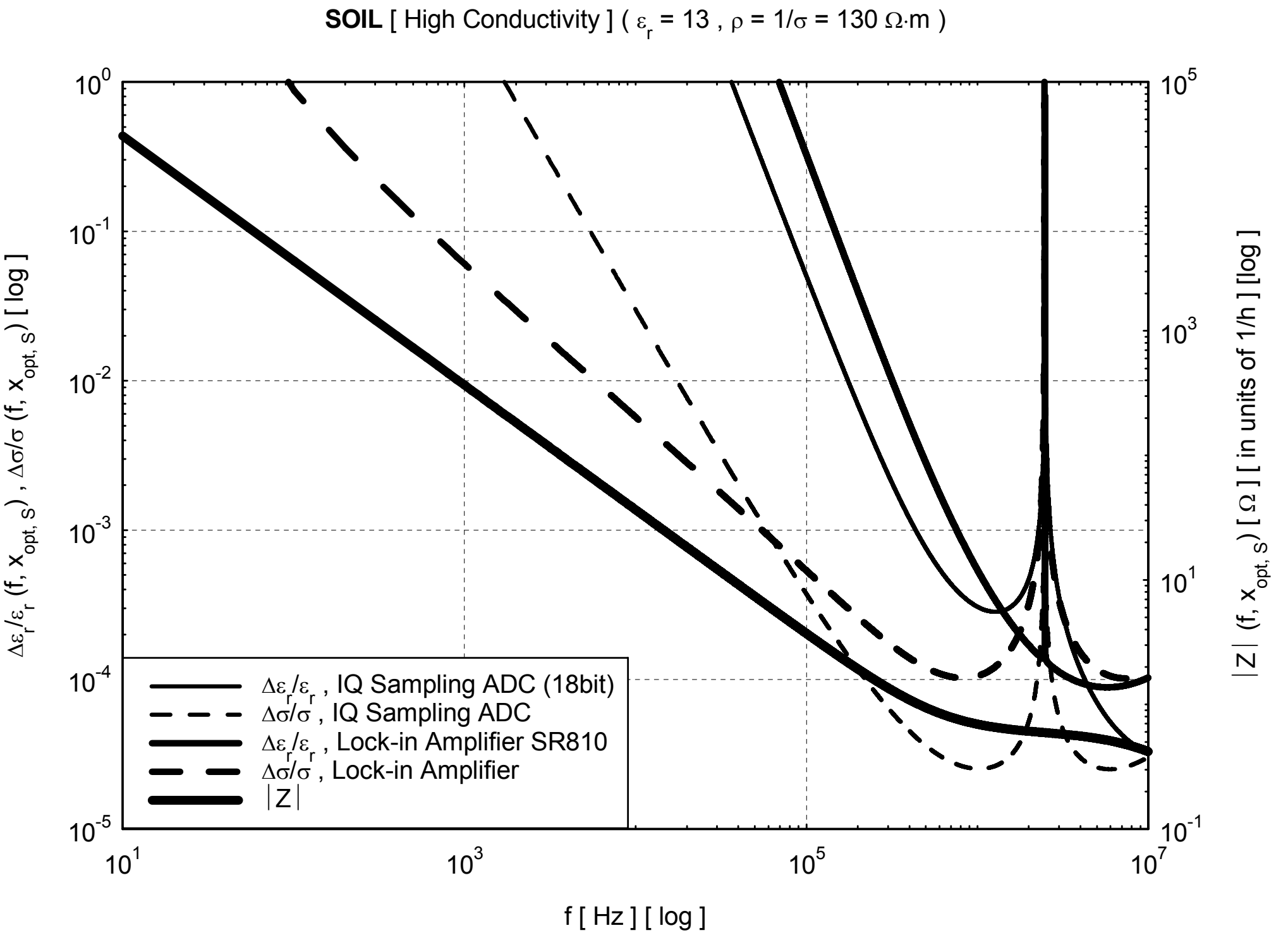

Figure 4.b

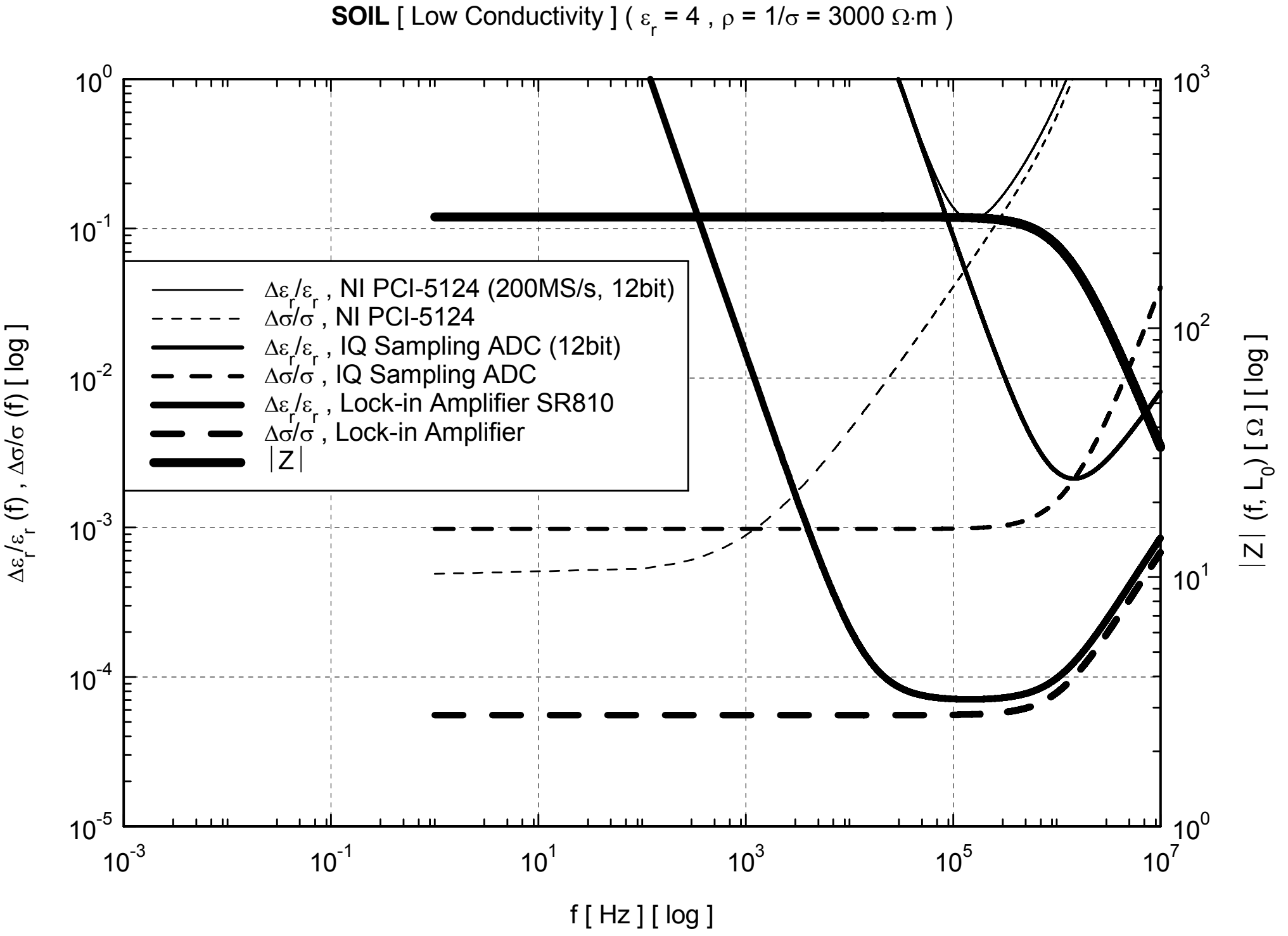

Figure 4.b.bis

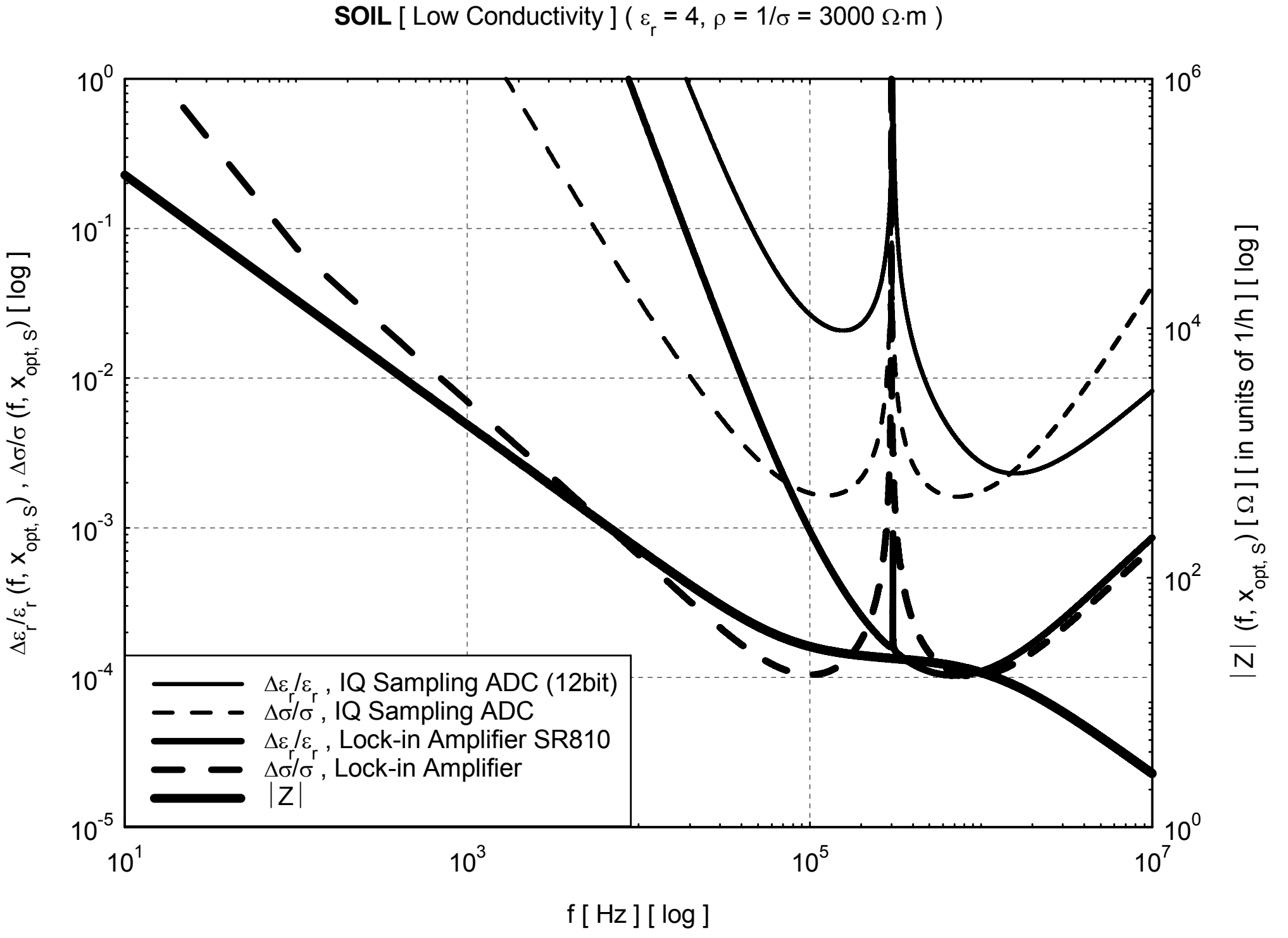

Figure 4.c

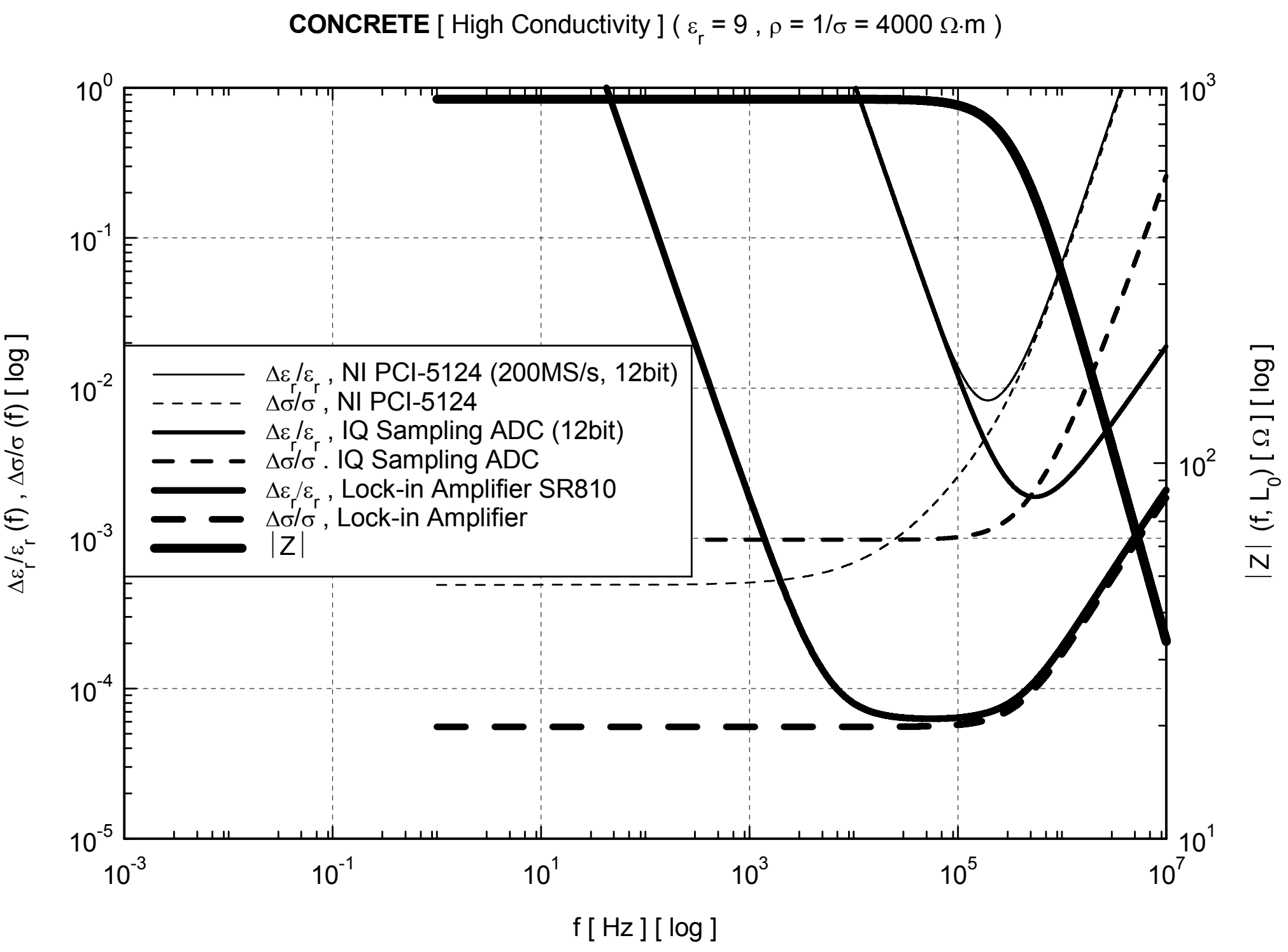

Figure 4.c.bis

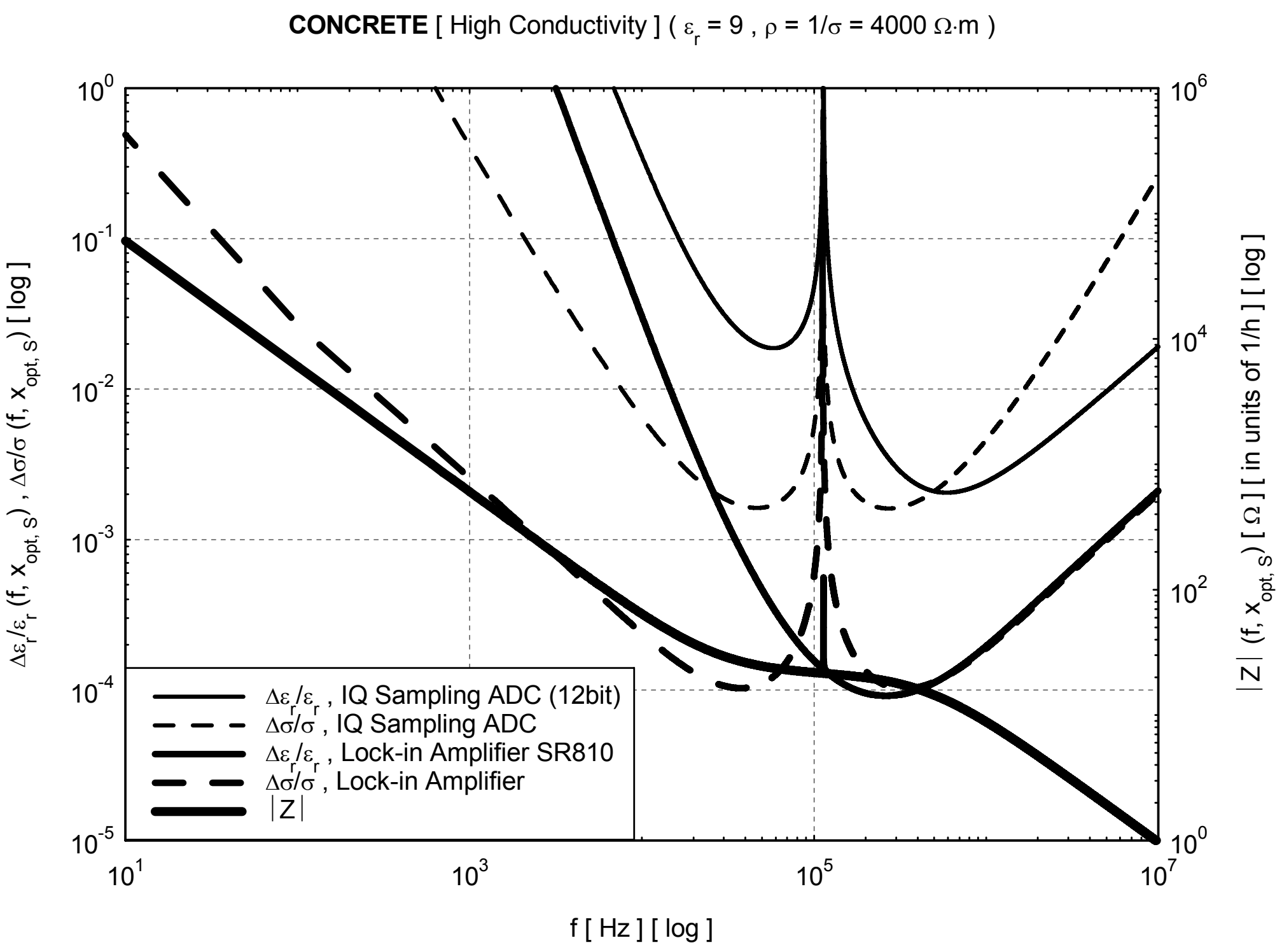

Figure 4.d

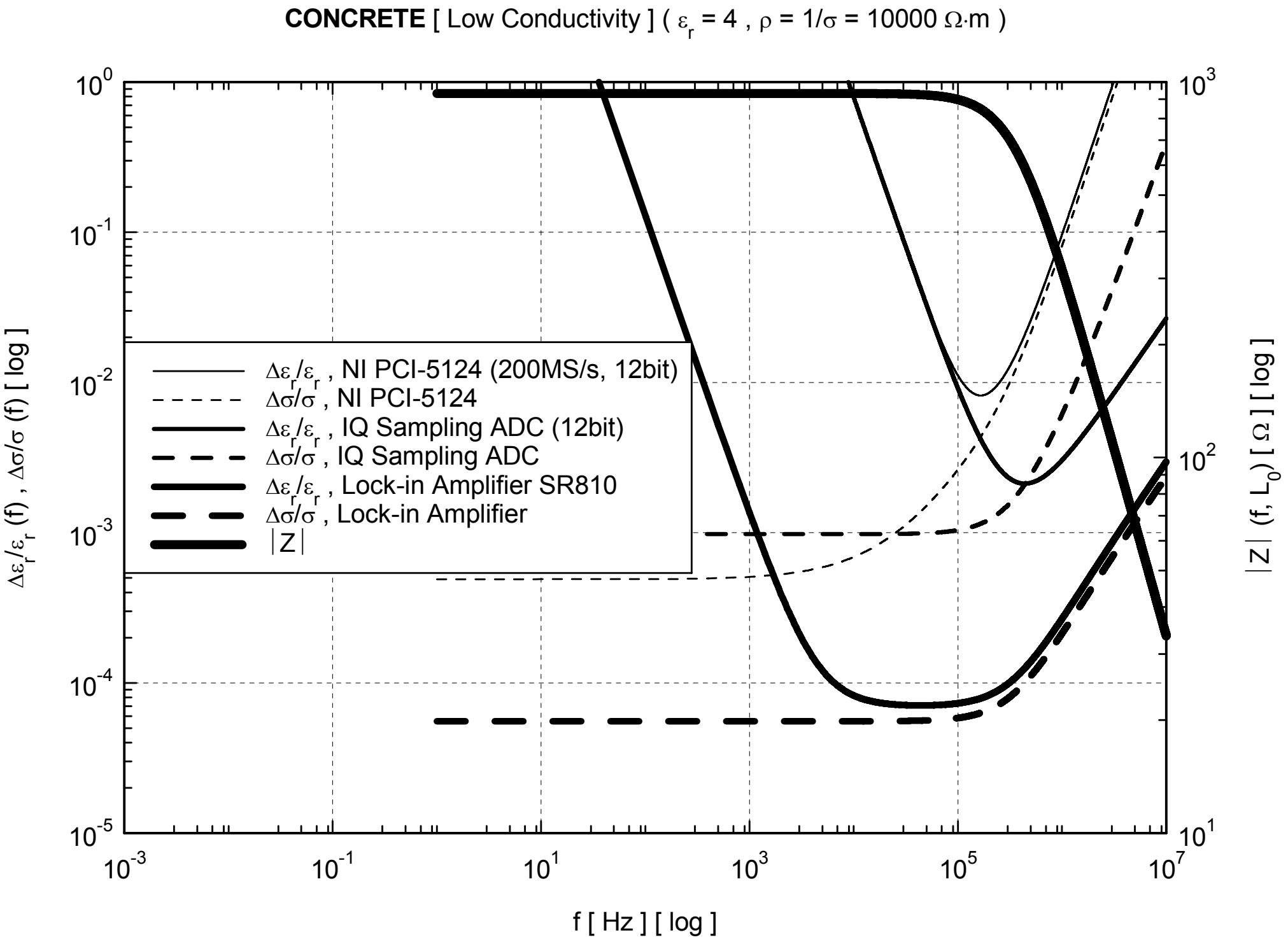

Figure 4.d.bis

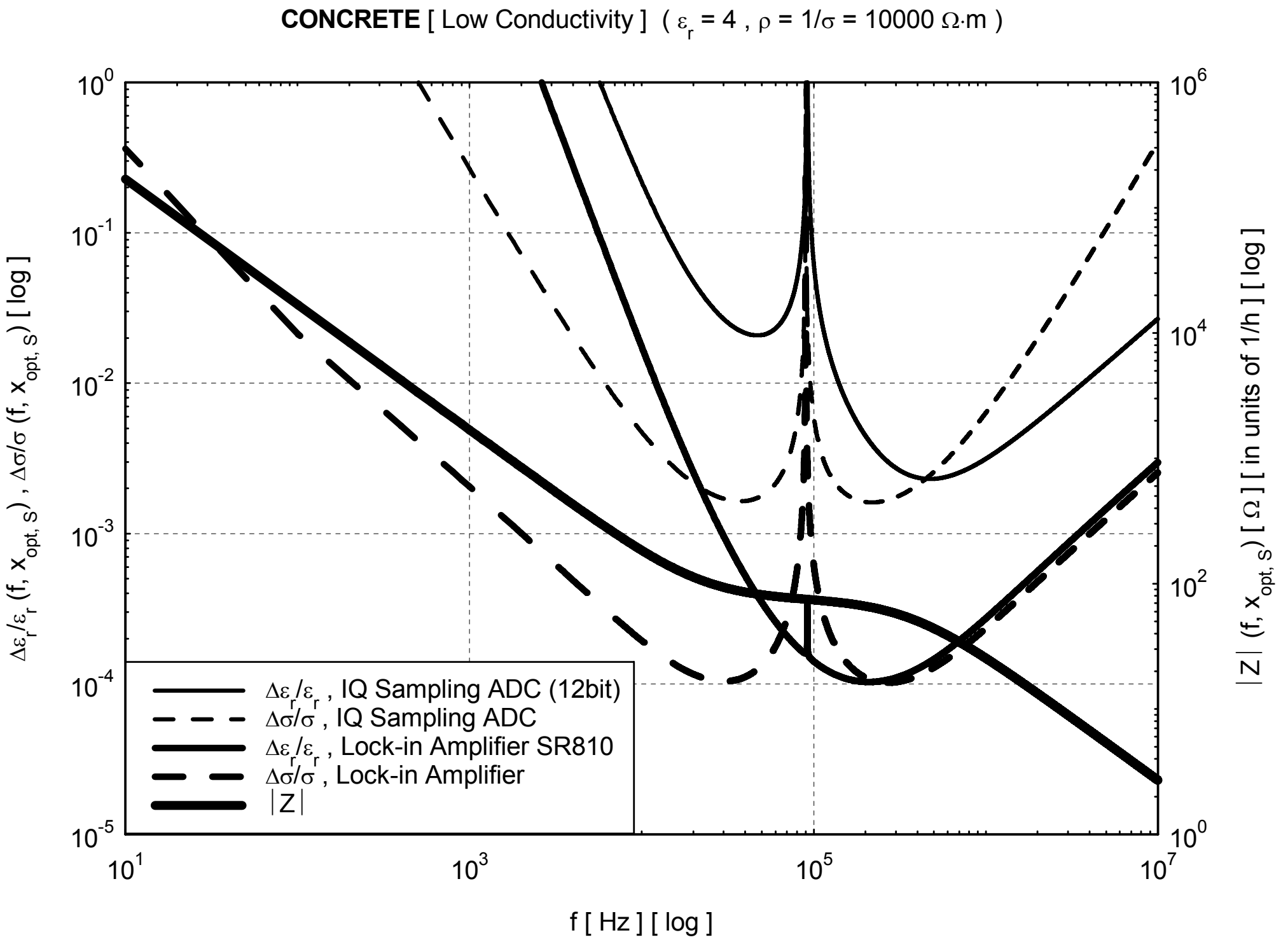

Figure 5.a

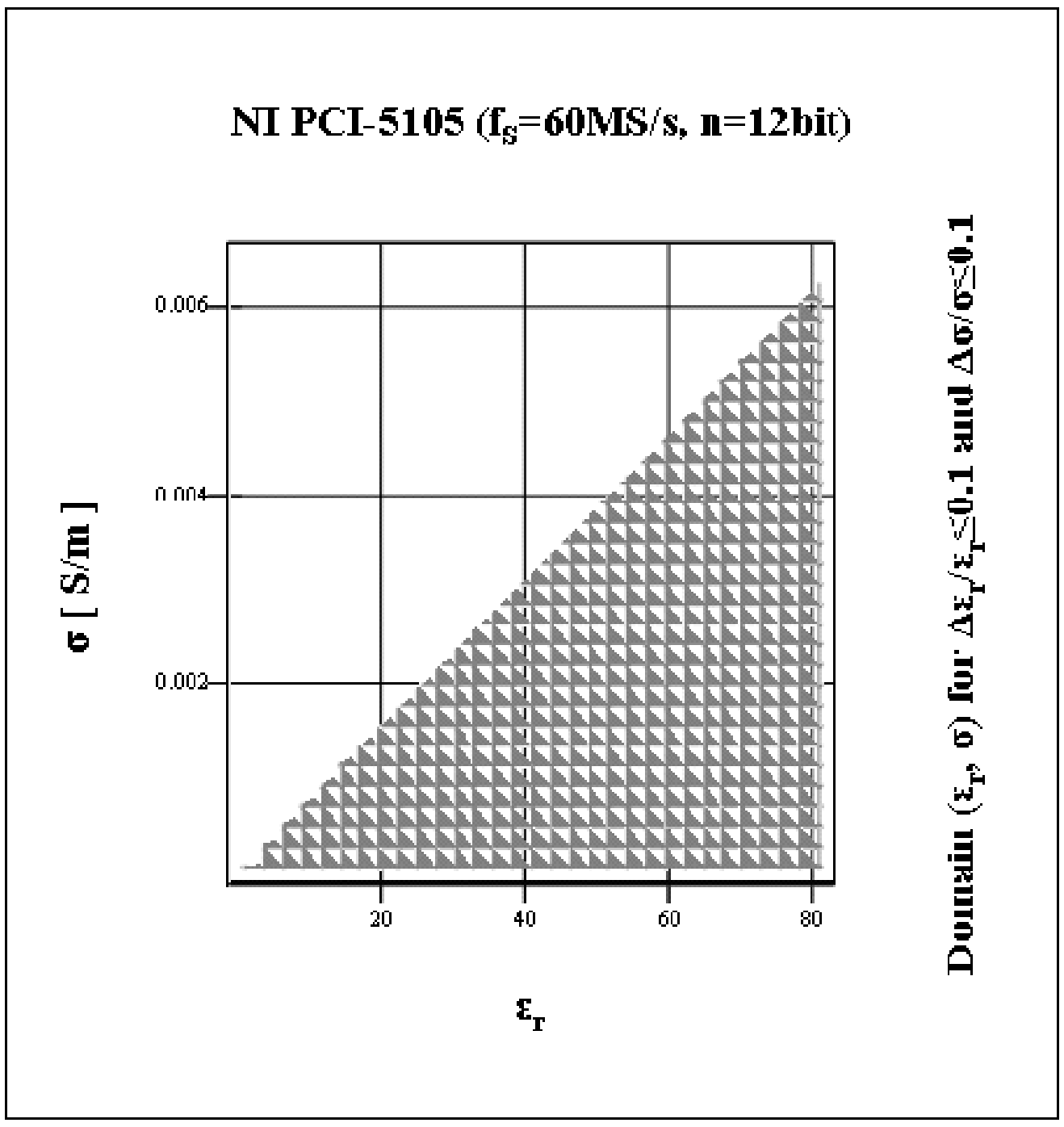

Figure 5.b

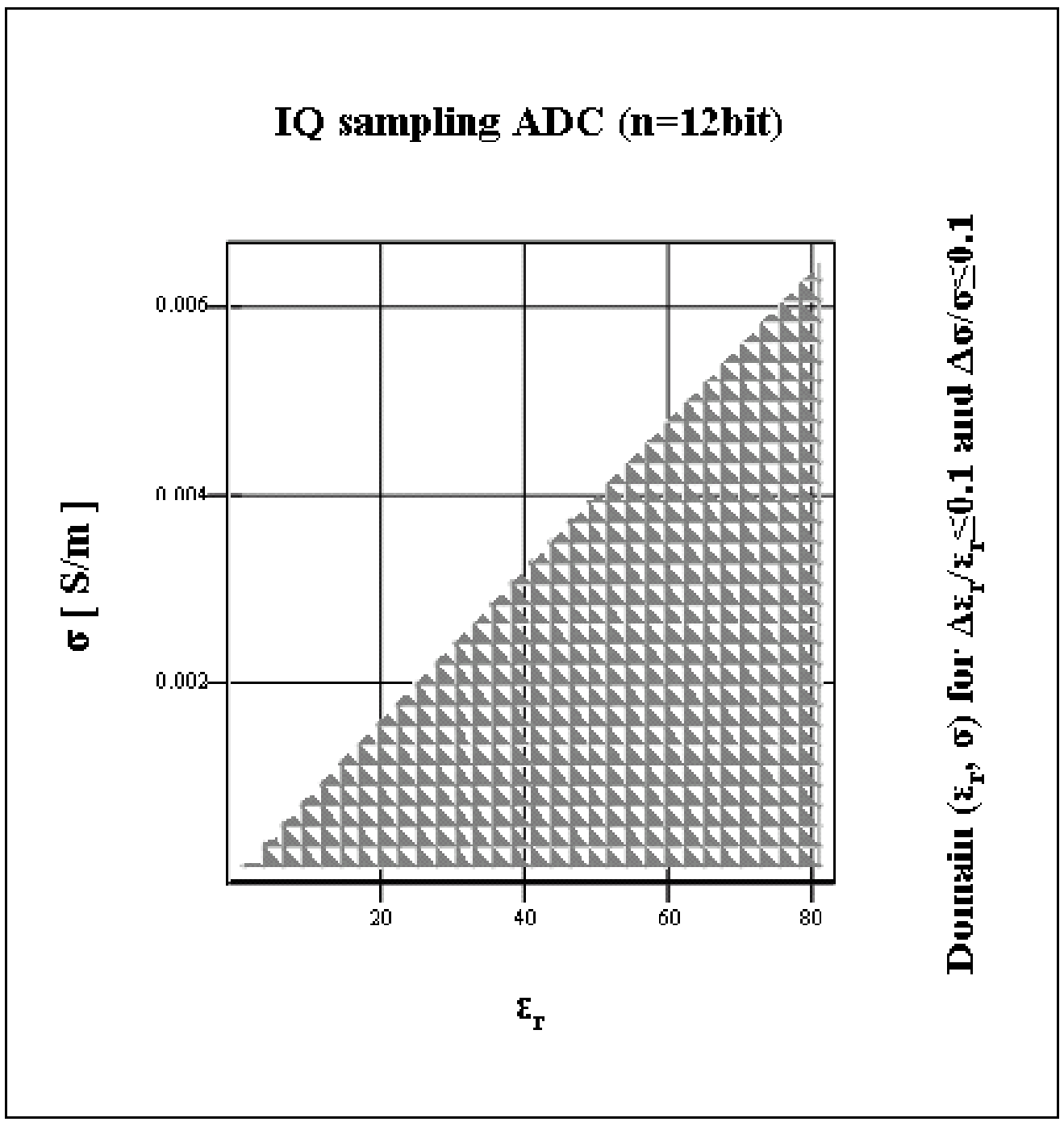

Figure 5.c

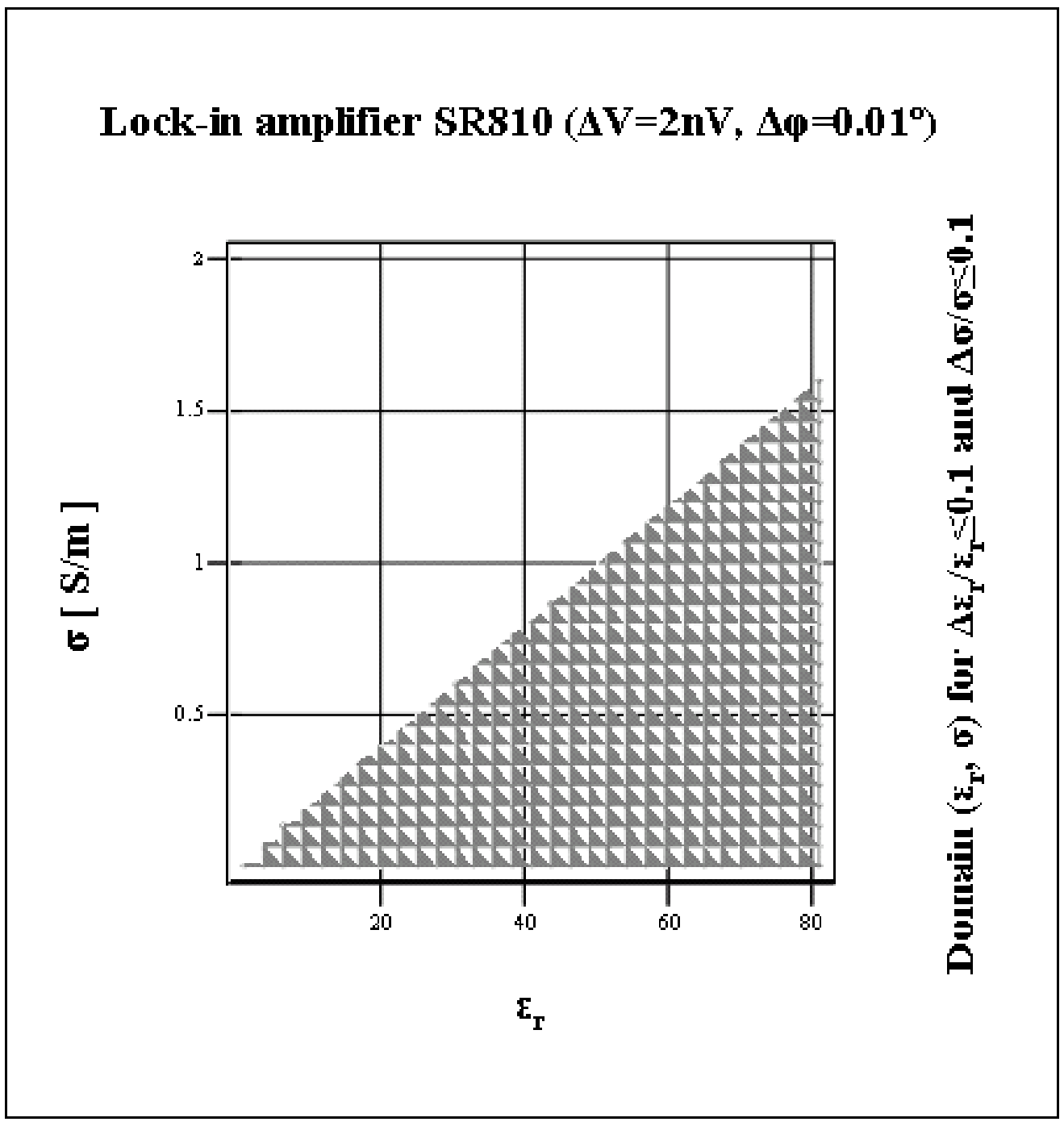
Lock-in amplifier SR810 (ΔV=2nV, Δφ=0.01°)
σ [ S/m ]
2
1.5
1
0.5
20
40
60
80
εr
Domain (εr, σ) for Δεr/εr≤0.1 and Δσ/σ≤0.1

Figure 6.a

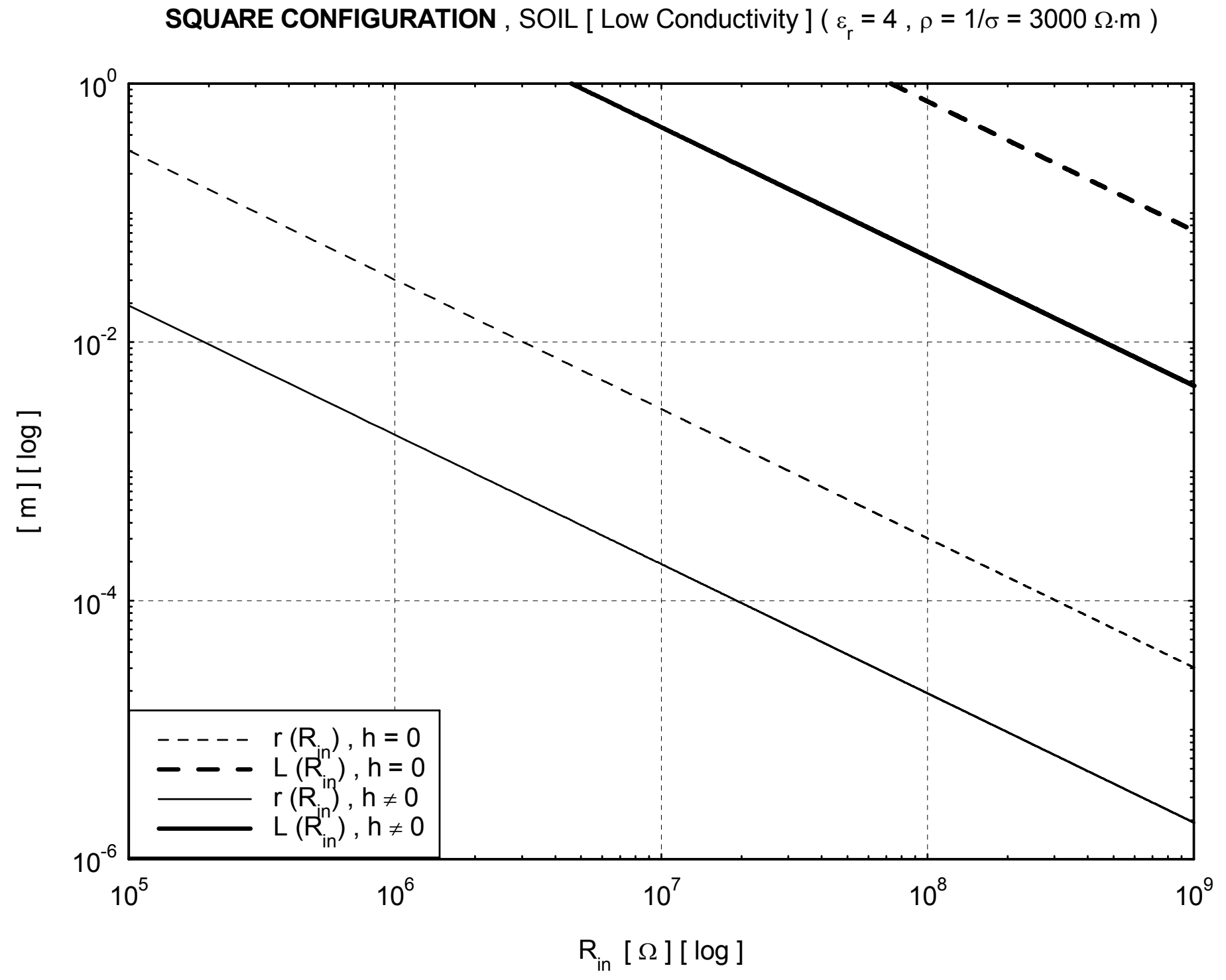

Figure 6.b

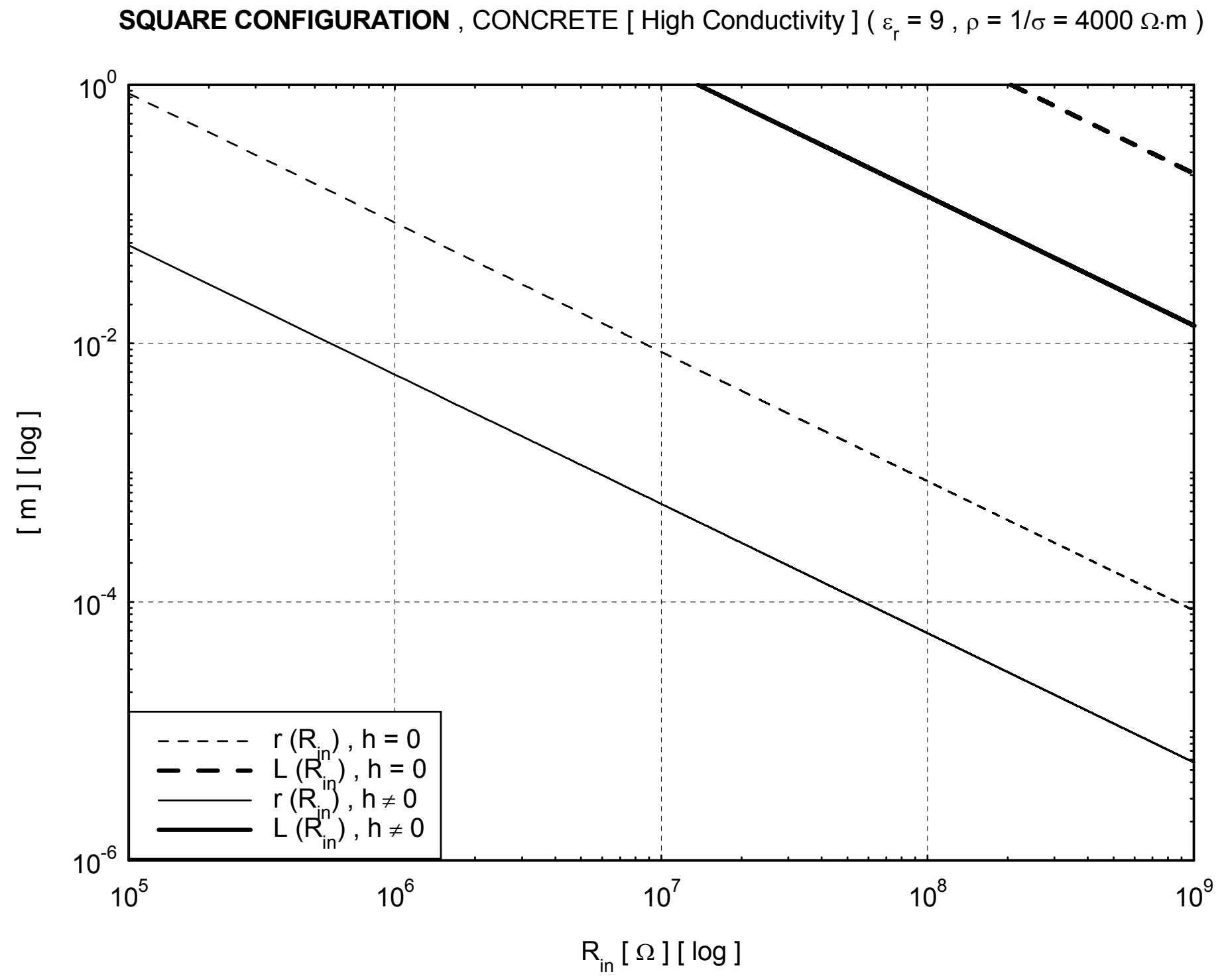

Figure 6.c

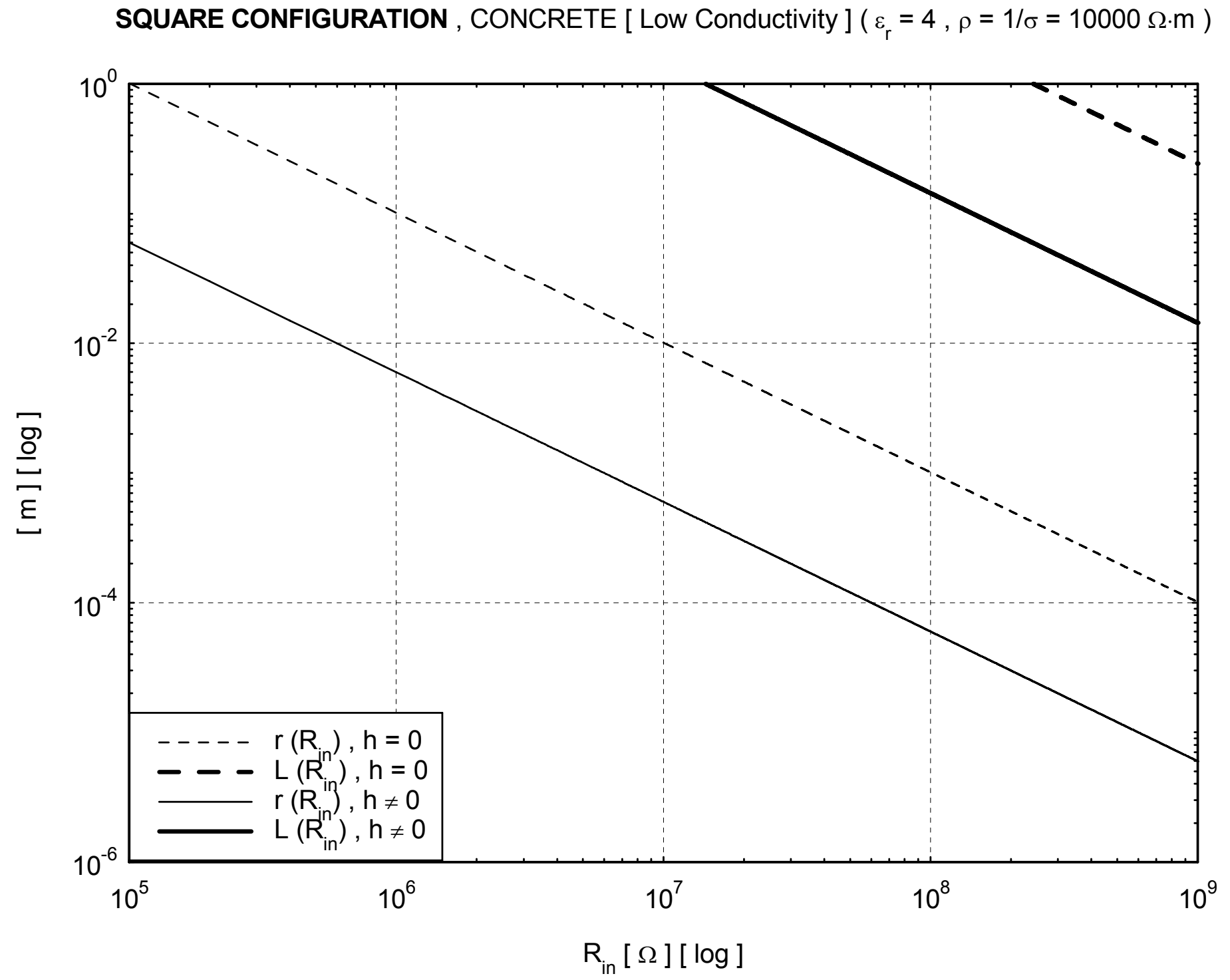

Figure 6.d

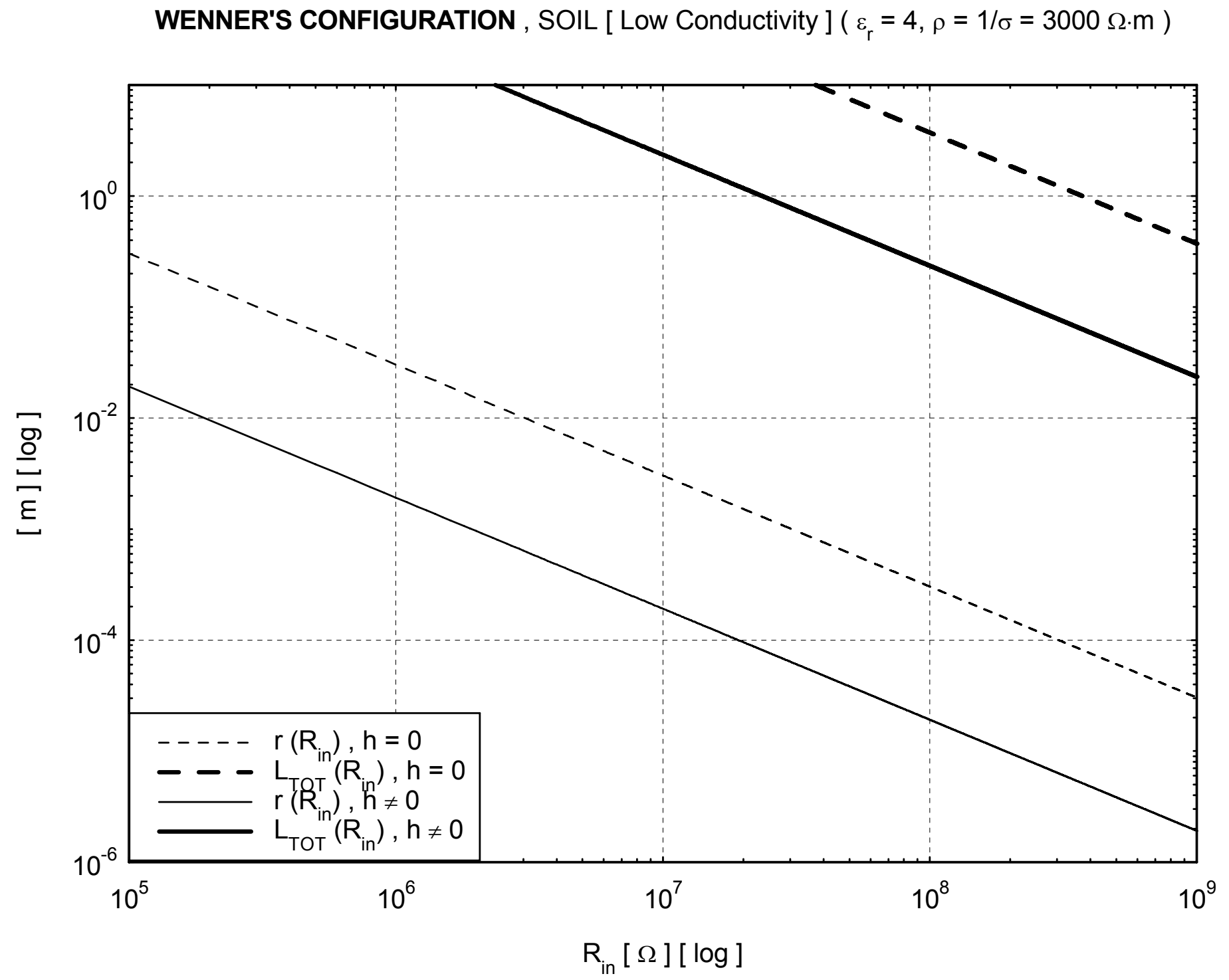

Figure 7.a

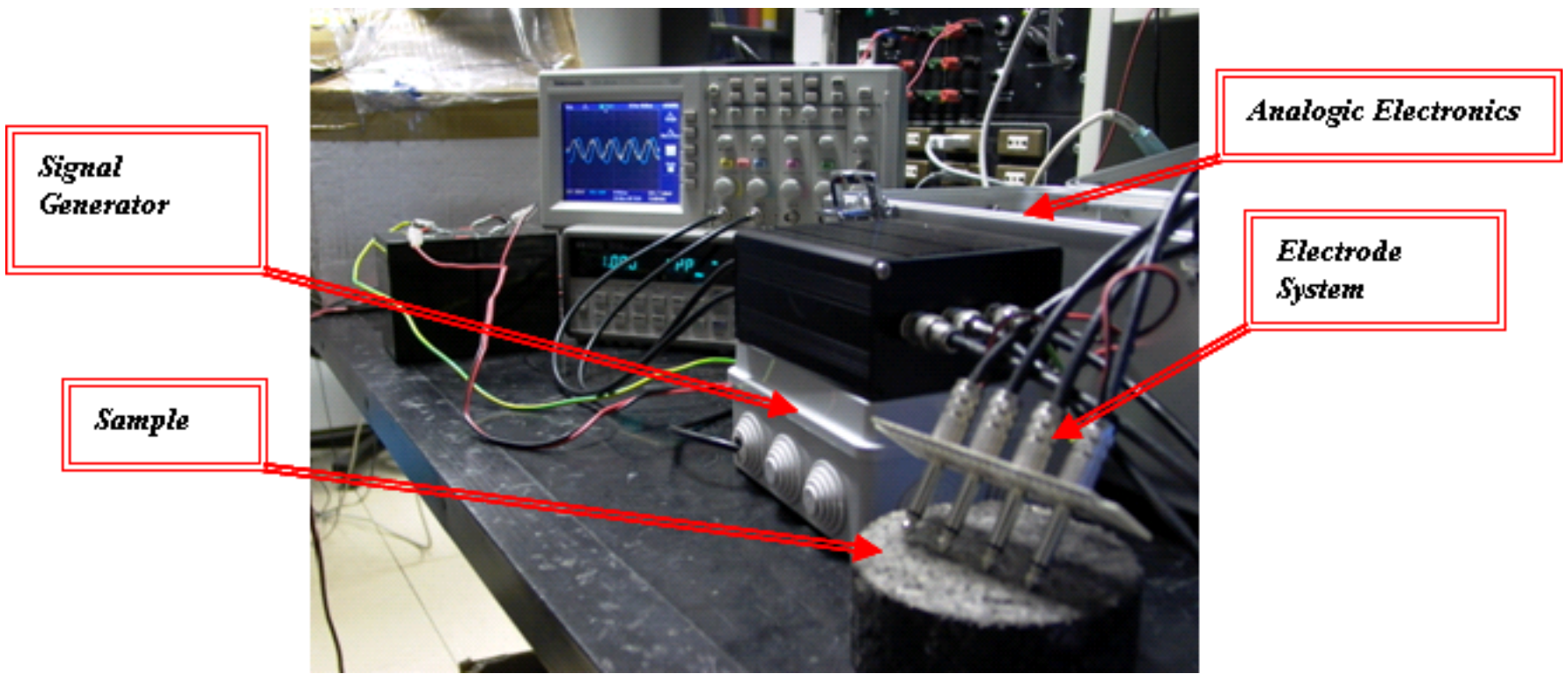

Figure 7.b

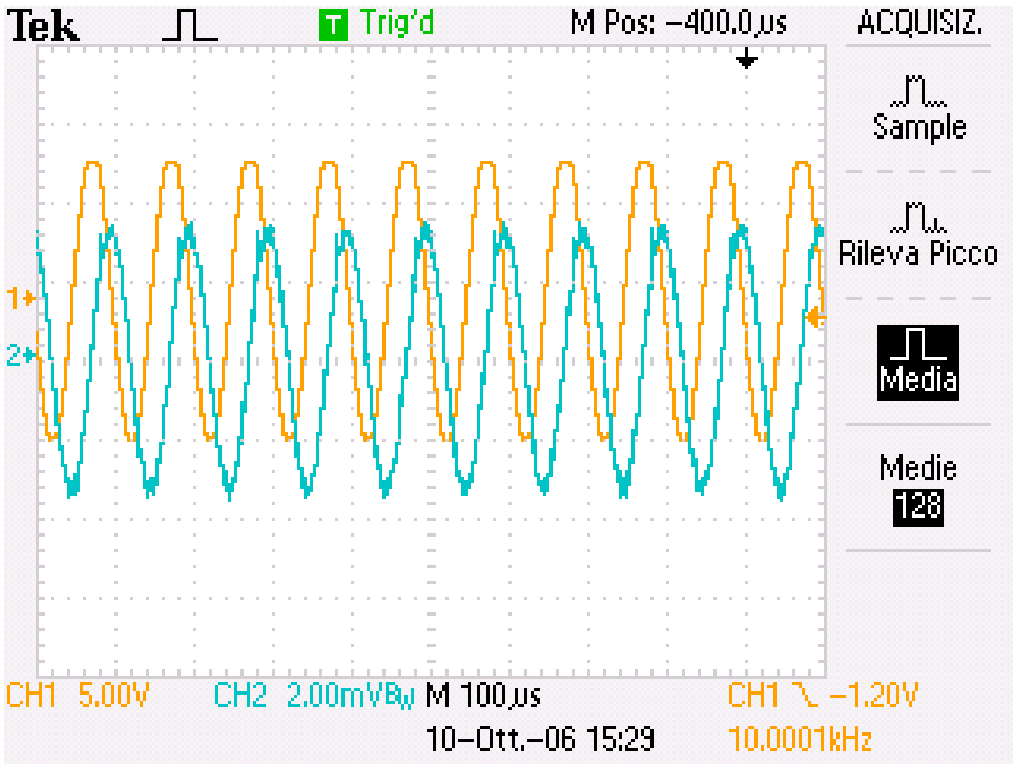

Table 1.a

| ***h = 0***<br>***SOIL [High Conductivity]***<br>***($\varepsilon_r$ = 13,***<br>***$\rho$ = 1/$\sigma$ =130 Ω·m)*** | **NI PCI-5922**<br>**($f_S$ = 10 MS/s,**<br>**n =18 bit)** | **IQ Sampling ADC**<br>**($n_{min}$ = 18)** | **Lock-in Amplifier SR810**<br>**(ΔV = 2nV,**<br>**Δφ = 0.01°)** |
|---|---|---|---|
| $f_{opt}$ | 154.935 kHz | 12.025 MHz | 1.199 MHz |
| $f_{min}$ | 83.903 kHz | 89.559 kHz | 2.9 kHz |
| $f_{max}$ | 330.926 kHz | 1.1236 GHz | 10.333 GHz |

Table 1.a.bis

| ***h ≠ 0***<br>***SOIL [High Conductivity]***<br>***($\varepsilon_r$ = 13,***<br>***$\rho$ = 1/$\sigma$ = 130 Ω·m)*** | **IQ Sampling ADC**<br>**($n_{min}$ = 18)** | **Lock-in Amplifier SR810**<br>**(ΔV = 2nV,**<br>**Δφ = 0.01°)** |
|---|---|---|
| $f_{up}$ | 1.024 MHz | 825.975 kHz |
| $f_{low}$ | 10.663 MHz | 22.402 MHz |

Table 1.b

| ***h = 0***<br>***SOIL [Low Conductivity]***<br>***($\varepsilon_r$ = 4,***<br>***$\rho$ = 1/$\sigma$ = 3000 Ω·m)*** | **NI PCI-5124**<br>**($f_S$ = 200 MS/s,**<br>**n = 12 bit)** | **IQ Sampling ADC**<br>**($n_{min}$ = 12)** | **Lock-in Amplifier SR810**<br>**(ΔV = 2nV,**<br>**Δφ = 0.01°)** |
|---|---|---|---|
| $f_{opt}$ | 387.772 kHz | 1.459 MHz | 145.489 kHz |
| $f_{min}$ | 95.055 kHz | 94.228 kHz | 379.08 Hz |
| $f_{max}$ | 1.832 MHz | 16.224 MHz | 1.099 GHz |

Table 1.b.bis

| ***h ≠ 0***<br>***SOIL [Low Conductivity]***<br>***($\varepsilon_r$ = 4,***<br>***$\rho$ = 1/$\sigma$ = 3000 Ω·m)*** | **IQ Sampling ADC**<br>**($n_{min}$ = 12)** | **Lock-in Amplifier SR810**<br>**(ΔV = 2nV,**<br>**Δφ = 0.01°)** |
|---|---|---|
| $f_{up}$ | 124.768 kHz | 100.508 kHz |
| $f_{low}$ | 1.492 MHz | 780.277 kHz |

Table 1.c

| ***h = 0*** ***CONCRETE*** ***[High Conductivity]*** ***($\varepsilon_r$ = 9,*** ***$\rho = 1/\sigma$ = 4000 Ω·m)*** | **NI PCI-5124 ($f_S$ = 200 MS/s, n = 12 bit)** | **IQ Sampling ADC (n = 12)** | **Lock-in Amplifier SR810 (ΔV = 2nV, Δφ = 0.01°)** |
|---|---|---|---|
| $f_{opt}$ | 193.94 kHz | 547.144 kHz | 54.558 kHz |
| $f_{min}$ | 33.315 kHz | 33.292 kHz | 134.02 Hz |
| $f_{max}$ | 1.206 MHz | 6.084 MHz | 463.74 MHz |

Table 1.c.bis

| ***h ≠ 0*** ***CONCRETE*** ***[High Conductivity]*** ***($\varepsilon_r$ = 9,*** ***$\rho = 1/\sigma$ = 4000 Ω·m)*** | **IQ Sampling ADC (n = 12)** | **Lock-in Amplifier SR810 (ΔV = 2nV, Δφ = 0.01°)** |
|---|---|---|
| $f_{up}$ | 46.648 kHz | 37.606 kHz |
| $f_{low}$ | 499.469 kHz | 546.487 kHz |

Table 1.d

| ***h = 0*** ***CONCRETE*** ***[Low Conductivity]*** ***($\varepsilon_r$ = 4,*** ***$\rho = 1/\sigma$ = 10000 Ω·m)*** | **NI PCI-5124 ($f_S$ = 200 MS/s, n = 12 bit)** | **IQ Sampling ADC (n = 12)** | **Lock-in Amplifier SR810 (ΔV = 2nV, Δφ = 0.01°)** |
|---|---|---|---|
| $f_{opt}$ | 165.329 kHz | 437.637 kHz | 43.647 kHz |
| $f_{min}$ | 28.273 kHz | 28.268 kHz | 113.724 Hz |
| $f_{max}$ | 1.016 MHz | 4.867 MHz | 329.796 MHz |

Table 1.d.bis

| ***h ≠ 0*** ***CONCRETE*** ***[Low Conductivity]*** ***($\varepsilon_r$ = 4,*** ***$\rho = 1/\sigma$ = 10000 Ω·m)*** | **IQ Sampling ADC (n = 12)** | **Lock-in Amplifier SR810 (ΔV = 2nV, Δφ = 0.01°)** |
|---|---|---|
| $f_{up}$ | 37.43 kHz | 30.152 kHz |
| $f_{low}$ | 477.58 kHz | 234.083 kHz |

Table 2.a

| ***SQUARE CONFIGURATION, SOIL [Low Conductivity] ($\varepsilon_r$ = 4, ρ = 1/σ = 3000 Ω·m)*** | **Galvanic Contact** | **Capacitive Contact** |
|---|---|---|
| *L* | ≈ 1 m | ≈ 1 m |
| $R_{in}$ | 72.84 MΩ | 4.6 MΩ |
| *r* | 416.813 µm | 416.836 µm |

Table 2.b

| ***SQUARE CONFIGURATION, CONCRETE [High Conductivity] ($\varepsilon_r$ = 9, ρ = 1/σ = 4000 Ω·m)*** | **Galvanic Contact** | **Capacitive Contact** |
|---|---|---|
| *L* | ≈ 1 m | ≈ 1 m |
| $R_{in}$ | 206.1 MΩ | 13.74 MΩ |
| *r* | 416.94 µm | 416.866 µm |

Table 2.c

| ***SQUARE CONFIGURATION, CONCRETE [Low Conductivity] ($\varepsilon_r$ = 4, ρ = 1/σ = 10000 Ω·m)*** | **Galvanic Contact** | **Capacitive Contact** |
|---|---|---|
| *L* | ≈ 1 m | ≈ 1 m |
| $R_{in}$ | 242.8 MΩ | 14.37 MΩ |
| *r* | 416.819 µm | 416.858 µm |

Table 2.d

| ***WENNER'S LINEAR CONFIGURATION, SOIL [Low Conductivity] ($\varepsilon_r$ = 4, ρ = 1/σ = 3000 Ω·m)*** | **Galvanic Contact** | **Capacitive Contact** |
|---|---|---|
| $L_{TOT} = 3 \cdot L$ | ≈ 10 m | ≈ 10 m |
| $R_{in}$ | 37.3 MΩ | 2.356 MΩ |
| *r* | 813.959 µm | 813.856 µm |

Table 3

| | **Electrical Conductivity measured by the prototype** | **Dielectric Permittivity measured by the prototype** |
|---|---|---|
| *Air* | 9.1 $10^{-8}$ S/m | 1.28 |
| *Coarse Grain Tar* | 5.0 $10^{-7}$ S/m | 6.27 |
| *Fine Grain Tar* | 3.54 $10^{-7}$ S/m | 4.33 |
| *Aggreagate Concrete* | 2.4 $10^{-6}$ S/m | 7.05 |

Fig. 1. Equivalent circuit of the RESPER.

Fig. 1. Circuito equivalente del RESPER.

Fig. 2. Electrical scheme for the analogical part of the measuring system: a signal generator (1), coupled to an amplifier stage, feeds one of two current electrodes (T1). The same current signal, picked to the other electrode (T2), is converted into voltage (2) and then amplified (3). The stage of differentiation for the voltage (4) is preceded by the feedback device to compensate the parasite capacities (5). The signal is sent to an analogical digital converter (ADC) and transferred to a personal computer, where it can be properly processed. The electronic circuit is composed primarily from two stages. The first consists of a current-voltage converter followed by a cascade of amplifiers, to amplify the weak currents typical of high impedances and the second consists of a voltage amplifier with a retroactive chain of capacitive compensation. The circuit has been designed to work linearly at LF in the band from DC to *100kHz*. The selected components have been developed specifically for electronic instruments of precision. The circuit techniques adopted for the compensation of the parasite capacities are innovative and allow to perform measurements of high impedances. This analogical device is connected to an analogical digital conversion board which contains even a digital analogical converter (DAC) used as a signal generator that, properly projected, can generate a whole series of measurements in an automatic way even at different frequencies for a full analysis.

Fig. 2. Schema elettrico per la parte analogica dello strumento: un generatore di segnale (1), accoppiato ad uno stadio di amplificazione, alimenta uno dei due elettrodi di corrente (T1). Lo stesso segnale di corrente, prelevato dall'altro elettrodo (T2), viene convertito in tensione (2) e poi amplificato (3). Lo stadio di differenziazione per la tensione (4) è preceduto dal dispositivo di controreazione per compensare le capacità parassite (5). Il segnale viene inviato ad un convertitore analogico digitale *(ADC)* e trasferito ad un *personal computer*, dove può essere opportunamente elaborato. Il circuito elettronico è composto principalmente da due stadi. Il primo è composto da un convertitore corrente tensione seguito da una cascata di amplificatori, per amplificare le deboli correnti caratteristiche di impedenze elevate e il secondo è composto da un amplificatore in tensione con una catena retroattiva di compensazione capacitiva. Il circuito è stato progettato per lavorare linearmente all'interno delle basse frequenze (*LF*) nella banda da *DC* a *100kHz.* I componenti selezionati sono stati realizzati appositamente per la strumentazione elettronica di precisione. Le tecniche circuitali adottate per la compensazione delle capacità parassite sono innovative e permettono di realizzare misure di impedenze elevate. Questo dispositivo analogico è collegato ad una scheda di conversione analogico digitale che contiene anche un convertitore digitale analogico (*DAC)* utilizzato come generatore di segnali che, programmato adeguatamente, permette di generare tutta una serie di misure in modo automatico anche a diverse frequenze per una analisi completa.

Figs. 3. RESPER in square (a) or linear (Wenner's) (b) configuration.

Figs.3. RESPER in configurazione a quadrato (a) o lineare (di Wenner)(b).

Figs. 4. Bode's diagrams for the inaccuracies $\Delta\varepsilon_r/\varepsilon_r$ and $\Delta\sigma/\sigma$ in the measurement of the dielectric permittivity $\varepsilon_r$ and the electrical conductivity $\sigma$, or for the modulus $|Z|$ of the transfer impedance, plotted as functions of the frequency $f$. The RESPER has a galvanic (height above ground $h=0$) (a-d) or capacitive (height/dimension ratio $x=x_{opt,S}$ optimally sized) (a.bis-d.bis) contact on non-saturated terrestrial soils characterized by an high ($\rho=1/\sigma=130\Omega\cdot m,\ \varepsilon_r=13$) (a, a.bis) and low ($\rho=3000\Omega\cdot m,\ \varepsilon_r=4$) (b, b.bis) conductivity or concretes with an high ($\rho=1/\sigma=4000\Omega\cdot m,\ \varepsilon_r=9$) (c, c.bis) and low ($\rho=10000\Omega\cdot m,\ \varepsilon_r=4$) (d, d.bis) conductivity. The quadrupolar probe is designed in the square configuration ($L_0=1m$) and is connected to a lock-in amplifier with specifications only similar to the SR810 and SR830's ones ($\Delta V=2nV,\ \Delta\varphi=0.01°$), commercialized by the Standford Research Systems Company, or a uniform sampling ADC only similar to the NI PCI-5922 $(f_S=10MS/s,\ n=18bit)$ and NI PCI-5124 $(f_S=200MS/s,\ n=12bit)$, commercialized by the National Instruments Company, or otherwise one of the IQ sampling ADCs with $n=18bit$ and $n=12bit$, which are being projected in our laboratories under the worst operative conditions, when an internal quartz is oscillating at its lowest merit factor $Q\approx10^4$. The ADCs allow the lowest minimum value of frequency $f_{min}$, within the band $B=100kHz$, such that the inaccuracies in the measurements result below a prefixed limit, *15%* referring to (a, a.bis) and *10%* for (b-d, b.bis-d.bis) [Tabs. 1].

Figs. 4. Diagrammi di Bode per le incertezze $\Delta\varepsilon_r/\varepsilon_r$ and $\Delta\sigma/\sigma$ nelle misure della permettività dielettrica $\varepsilon_r$ e conducibilità elettrica $\sigma$, o per il modulo $|Z|$ della impedenza di trasferimento, tracciate come funzioni della frequenza $f$. Il RESPER presenta un contatto galvanico (altezza da terra $h=0$) (a-d) o capacitivo (rapporto altezza/dimensione $x=x_{opt,S}$, calibrato ottimamente) (a.bis-d.bis) su suoli terrestri non-saturi caratterizzati da un alta ($\rho=1/\sigma=130\Omega\cdot m,\ \varepsilon_r=13$) (a, a.bis) e bassa ($\rho=3000\Omega\cdot m,\ \varepsilon_r=4$) (b, b.bis) conducibilità o calcestruzzi non saturi con un alta ($\rho=1/\sigma=4000\Omega\cdot m,\ \varepsilon_r=9$) (c, c.bis) e bassa ($\rho=10000\Omega\cdot m,\ \varepsilon_r=4$) (d, d.bis) conducibilità. La sonda a quadripolo è progettata nella configurazione a quadrato ($L_0=1m$) ed è connessa ad un amplificatore *lock-in* con specifiche solo simili a quelle dei SR810 e SR830 ($\Delta V=2nV,\ \Delta\varphi=0.01°$), commercializzati dalla Standford Research Systems Company, o un ADC a campionamento uniforme solo simile ai NI PCI-5922 $(f_S=10MS/s,\ n=18bit)$ e NI PCI-5124 $(f_S=200MS/s,\ n=12bit)$, commercializzati dalla National Instruments Company, o altrimenti uno degli ADC a campionamento IQ con $n=18bit$ e $n=12bit$, che sono in fase di progettazione nei nostri laboratori in condizioni peggiorative, quando un quarzo interno oscilla al suo più basso fattore di merito $Q\approx10^4$. L'ADC consente il più basso valore minimo di frequenza $f_{min}$, entro la banda $B=100kHz$, in modo tale che le incertezze nelle misure risultino al di sotto di un limite prefissato, il *15%* con riferimento a (a, a.bis) e il *10%* per (b-d, b.bis-d.bis) [Tab. 1].

Figs. 5. Plots for the domain $(\sigma,\varepsilon_r)$ of the electrical conductivity $\sigma$ and the dielectric permittivity $\varepsilon_r$ such that both the inaccuracies $\Delta\sigma/\sigma(\sigma,\varepsilon_r)$, in the measurement of the conductivity $\sigma$, and $\Delta\varepsilon_r/\varepsilon_r(\sigma,\varepsilon_r)$, of the permittivity $\varepsilon_r$, result below a prefixed limit of *10%.* The RESPER ($B=100kHz$) has a galvanic contact with the subjacent medium ($h=0$) and is connected to a low-cost uniform sampling ADC with specifications similar to the NI PCI-5105's one $(f_S=60MS/s,\ n=12bit)$ (a), or an IQ sampling ADC, specified by $n=12bit$ (b), or otherwise a lock-in amplifier similar to the SR810 and SR830 ($\Delta V=2nV,\ \Delta\varphi=0.01°$) (c).

Figs. 5. Grafici per il dominio $(\sigma,\varepsilon_r)$ delle conducibilità elettrica $\sigma$ e permettività dielettrica $\varepsilon_r$ tale che sia le incertezze $\Delta\sigma/\sigma(\sigma,\varepsilon_r)$, nella misura della conducibilità $\sigma$, che $\Delta\varepsilon_r/\varepsilon_r(\sigma,\varepsilon_r)$, della permettività $\varepsilon_r$, risultino al di sotto di un limite prefissato del *10%.* Il RESPER ($B=100kHz$) presenta un contatto galvanico con il mezzo sottostante ($h=0$) ed è connesso ad un ADC a campionamento uniforme e basso costo con specifiche simili a quelle del NI PCI-5105 $(f_S=60MS/s,\ n=12bit)$ (a), o un ADC a campionamento IQ, specificato da $n=12bit$ (b), o altrimenti un amplificatore *lock-in* simile ai SR810 e SR830 ($\Delta V=2nV,\ \Delta\varphi=0.01°$) (c).

Figs. 6. Like-Bode's diagrams of the minimal radius $r(R_{in},f_{min})$ for the RESPER electrodes and of the characteristic geometrical dimension $L_S(r,n_{min})$ for the square configuration or the length $L_{TOT}(r,n_{min})=3\cdot L_W(r,n_{min})$ for the (Wenner's) linear configurations, plotted as functions of the input resistance $R_{in}$ for the amplifier stage. The quadrupole probe could be designed for both galvanic and capacitive contact: the square configuration, on non-saturated terrestrial soils characterized by a low conductivity ($\rho=3000\Omega\cdot m,\ \varepsilon_r=4$) (a) or concretes with an high ($\rho=1/\sigma=4000\Omega\cdot m,\ \varepsilon_r=9$) (b) and low ($\rho=10000\Omega\cdot m,\ \varepsilon_r=4$) (c) conductivity; the Wenner's linear configuration, for galvanic and capacitive contact on terrestrial soils characterized by a low conductivity ($\rho=3000\Omega\cdot m,\ \varepsilon_r=4$) (d) [Tabs. 2]. The quadrupole is connected to an IQ sampling ADC with minimal bit resolution $n_{min}=12$, which allows inaccuracies in the measurements below a prefixed limit (*10%*) within the LF band $[f_{min},\ f_{max}]$ for operative conditions of galvanic contact or in the MF-HF band $[f_{up},\ f_{low}]$ for capacitive contact [Tab. 1].

Figs. 6. Simil-diagrammi di Bode del raggio minimo $r(R_{in},f_{min})$ per gli elettrodi del RESPER e della dimensione geometrica caratteristica $L_S(r,n_{min})$ per la configurazione a quadrato o della lunghezza $L_{TOT}(r,n_{min})=3\cdot L_W(r,n_{min})$ per la configurazione lineare (o di Wenner), tracciati come funzioni della resistenza

di ingresso $R_{in}$ per lo stadio amplificatore. La sonda a quadripolo potrebbe essere progettata sia per il contatto galvanico che capacitivo: la configurazione a quadrato, su suoli terrestri non saturi caratterizzati da una bassa conducibilità ($\rho=3000\Omega\cdot m,\ \varepsilon_r=4$) (a) o calcestruzzi non saturi con un elevata ($\rho=1/\sigma=4000\Omega\cdot m,\ \varepsilon_r=9$) (b) e bassa ($\rho=10000\Omega\cdot m,\ \varepsilon_r=4$) (c) conducibilità; la configurazione lineare di Wenner, per contatto galvanico e capacitivo su suoli terrestri caratterizzati da una bassa conducibilità ($\rho=3000\Omega\cdot m,\ \varepsilon_r=4$) (d) [Tabs. 2]. Il quadripolo è connesso ad un ADC a campionamento IQ con risoluzione minima di bit $n_{min}=12$, che permette incertezze nelle misure al di sotto di un limite prefissato (*10%*) all'interno della banda LF $[f_{min}, f_{max}]$ per le condizioni operative di contatto galvanico o nella banda MF-HF $[f_{up}, f_{low}]$ per contatto capacitivo [Tab. 1].

Fig. 7.a. The first prototype of a measuring system during the test on a block of tar. One can see the high-voltage signal generator (white box), the analogical electronics (black box), the prototype system of electrodes, put on a sample of tar, and the oscilloscope which digitizes the two signals, current and voltage, and displays them in the time domain. In the future developing, the electronics will be further integrated to render the system as compact as possible.

Fig. 7.a. Il primo prototipo dello strumento durante la fase di test su di un blocco di catrame. Sono visibili il generatore di alta tensione (scatola bianca), l'elettronica analogica (scatola nera) e il sistema di punte prototipale (poggiato su un campione di catrame) e l'oscilloscopio, che digitalizza i due segnali, di corrente e di tensione, e li visualizza nel dominio del tempo. Nello sviluppo successivo, l'elettronica verrà ulteriormente integrata per rendere lo strumento il più compatto possibile.

Fig. 7.b. The two signals, of voltage $V$ (yellow line) and current $V_I$ (in blue), acquired on the sample of tar in fig, 7.a: the two signals are almost out of phase ($\sim 90°$), as expected for a predominantly capacitive transfer impedance.

Fig. 7.b I due segnali, di tensione $V$ (in giallo) e corrente $V_I$ (in blu), acquisiti sul campione di catrame in fig. 7.a: i due segnali sono quasi sfasati di ($\sim 90°$), come ci si aspetta per un'impedenza di trasferimento prevalentemente capacitiva.

Tab. 3. First results of laboratory tests, performed on samples of different materials. The values are compatible with those in literature. These initial steps have enabled to evaluate the validity of the instrumentation and the correctness of the used method. The device is shown stable varying the peak-peak amplitude of the signal for a frequency range which extends from few *kHz* to around *100kHz.*

Tab. 3. I primi risultati degli esami di laboratorio, effettuati su campioni di materiali diversi. I valori sono compatibili con quelli in letteratura. Questi passi iniziali hanno consentito di valutare la validità della strumentazione e la correttezza del metodo utilizzato. Il dispositivo si è dimostrato stabile, variando l'ampiezza di picco-picco del segnale per una banda di frequenze che si estende da pochi *kHz* a circa *100kHz*.